\documentclass[10pt,twocolumn,letterpaper]{article}

\newif\ifarxiv
\arxivtrue

\ifarxiv
  \usepackage[pagenumbers]{cvpr}   
\else
  \usepackage[review]{cvpr}        
\fi
\usepackage{verbatim}
\usepackage{tabularx,booktabs}

\usepackage{cuted} 

\usepackage{currfile} 

\usepackage{caption} 
\definecolor{cvprblue}{rgb}{0.21,0.49,0.74}
\usepackage[pagebackref,breaklinks,colorlinks,allcolors=cvprblue]{hyperref}
\usepackage{multirow}

\def\paperID{42431} 
\def\confName{CVPR}
\def\confYear{2027}

\title{Beyond Outlier Rejection: Transient-Aware Reconstruction of Multi-Stack MRI}

\author{%
\parbox{0.9\textwidth}{\centering
Anika Knupfer$^{1,2,3}$, Maximilian Lindholz$^{4,5}$, Maria Camila Bustos-Vivas$^{1,6}$, Joseph V. Hajnal$^{7}$, Bodo Rosenhahn$^{1,2}$, Jana Hutter$^{1,2,3,7}$\\[1.5ex]
{\small
$^{1}$Institute of Information Processing, Leibniz University Hannover, Hannover, Germany\\
$^{2}$Medizinische Hochschule Hannover, Hannover, Germany\\
$^{3}$CAIMED, L3S, Hannover, Germany\\
$^{4}$Department of Radiology, Charité Universitätsmedizin Berlin, Berlin, Germany\\
$^{5}$Berlin Institute of Health, Berlin, Germany\\
$^{6}$Institute for Radiology, University Hospital Erlangen, Erlangen, Germany\\
$^{7}$Biomedical Engineering Department, School of Biomedical Engineering and Imaging Sciences, KCL, London, UK\\[1ex]
{\tt knupfer@tnt.uni-hannover.de}}
}}

\begin{document}
\maketitle
\begin{abstract}
Clinical MRI typically acquires the organ of interest as consecutive stacks of 2D slices in different orientations. Reconstruction methods such as SVRTK and NeSVoR fuse them into one volume and down-weight what does not fit, although in living tissue the change during the scan is often the signal itself and can contain clinically relevant physiological information, such as fetal motion and uterine contraction. We propose an analysis-by-synthesis decomposition that explains every acquired voxel through its exact acquisition geometry as a static anatomy shared by all stacks, a rigid and smooth non-rigid motion per stack, and a sparse transient field per stack for changes during one acquisition only. A robust reweighting keeps such changes out of the anatomy. The model is fitted per subject without training data, with one configuration for all datasets. On a digital phantom with known ground truth, static changes are kept entirely in the anatomy, and the transient field of the stack that saw a transient recovers 68\% of it at 12\,mm radius. From identical inputs, the anatomy predicts held-out stacks better than SVRTK and NeSVoR on uterine (NCC 0.897 against 0.874 and 0.863, 123 patients) and on prostate MRI from unseen sites and scanners (0.847 against 0.795 and 0.757, 85 patients), and on fetal brain MRI as well as SVRTK and better than NeSVoR (0.771 against 0.763 and 0.747, 83 fetuses). The transient fields flag corrupted fetal slices with an AUROC of 0.94 against expert raters. As the input of a frozen anomaly detector trained on healthy subjects only, the anatomy localizes radiologist-delineated uterine lesions better than the SVRTK and NeSVoR volumes (AUROC 0.76 against 0.71 and 0.71, 37 patients).

\end{abstract} 
    
\section{Introduction} 
\label{sec:intro}
Obtaining insights into 3D objects from sparse, noisy and artifacted measurements acquired from different angles and conditions is a common, recurring problem across the disciplines~\cite{zou2024sparse3d}, from video observation~\cite{javadi2026video}, geophysics~\cite{loris2010nonlinear}, remote sensing~\cite{wang2024res}, non-destructive testing~\cite{duba2025plug,ahmad2026compensating}, archaeology~\cite{muratbekuly2025research} to medical imaging~\cite{sarmah2023survey}. Typical 3D reconstruction techniques assume the object to be static~\cite{wang2026learning} and focus on down-weighting or eliminating inconsistent measurements~\cite{uus2022automated}, acquisition optimization especially in clinical medicine often targets motion-free images by making life comply to the demands of imaging~\cite{feng2026respiratory}. The observable transient changes, however, carry crucial information themselves: fetal motion informs on neurological status~\cite{moreira2018clinical}, contractile activity reflects muscular tissue health~\cite{beausejour2024innovations} and peristaltic motion is a key biomarker for diagnosis and treatment monitoring in \eg bowel disease and in-vivo fertilization~\cite{hameed2025esr,elad2020biomechanics}.
\begin{figure}[t]
\centering
\includegraphics[width=0.8\columnwidth]{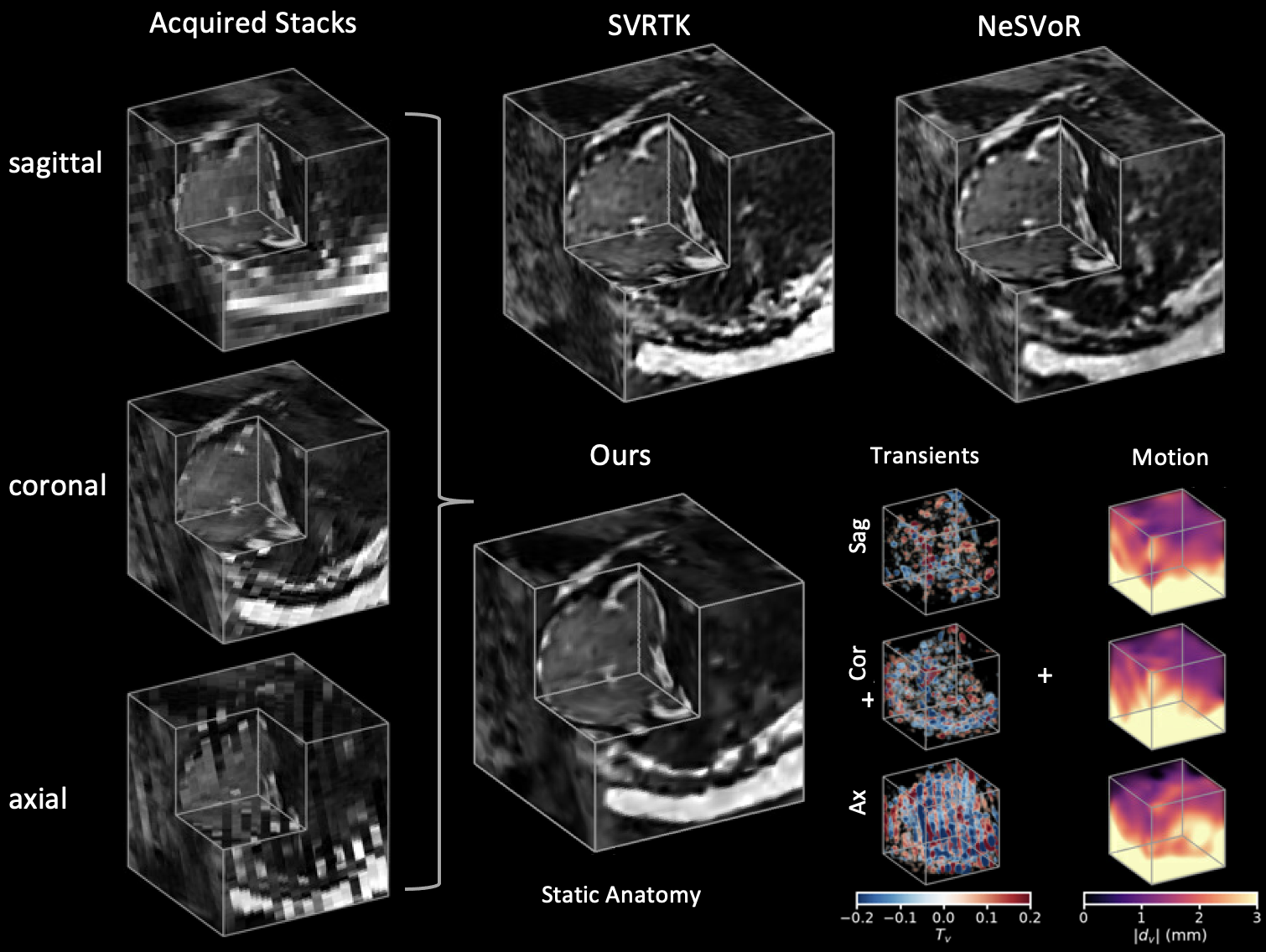}
\caption{Fetal brain MRI stacks (left). Slice-to-volume reconstruction (SVRTK~\cite{uus2020deformable}, NeSVoR~\cite{xu2023nesvor}, top) fuses them into one volume and discards what they disagree on. Our decomposition (bottom) recovers the static anatomy $A$ and keeps the disagreement as a transient field $T_v$ and non-rigid motion per stack.}
\label{fig:teaser}
\end{figure} 

Radiological data covers the area of interest in multiple views or "stacks" of 2D slices in sagittal, coronal and axial orientation, allowing the radiologist to resolve inconsistencies. In areas with substantial motion, with fetal MRI as a prime example, the individual slices are acquired fast to limit motion effects and slice-to-volume-reconstruction (SVR) approaches are used to fuse them into high resolution 3D representations (see Figure~\ref{fig:teaser}). Inconsistencies between slices arise from motion, transient physiology such as uterine contractions, characterized as non-rigid, transient changes in shape and signal, and artifacts such as dropouts and aliasing. Beyond medicine, MRI is also used to image porous media or non-metallic materials such as polymers and fibre-reinforced composites~\cite{coussot2020progress,alves2021detection}, where stacks of slices acquired over time face similar inconsistencies. Our experiments, however, are confined to medical data. Current SVR methods (Kuklisova-Murgasova~\cite{kuklisova2012reconstruction}, NiftyMIC~\cite{ebner2020automated}, SVRTK~\cite{uus2020deformable}, NeSVoR~\cite{xu2023nesvor}) down-weight or reject inconsistent slices as outliers. Outlier rejection is thereby often binary and unexplained, not distinguishing why a slice doesn’t fit (rigid motion, local deformation, artifact or a physiological event). Comparable vision methods (NeRF in the Wild~\cite{Martin-Brualla_2021_CVPR}, RobustNeRF~\cite{Sabour_2023_CVPR}, HybridGS~\cite{lin2024hybridgs}) also handle transients as distractors to remove, not as phenomena to characterize. On the other hand, dedicated dynamic acquisitions are increasingly added to assess physiology, in turn however not contributing to the anatomical information available. Our approach treats both in a unified framework, reinterpreting a limitation of anatomical MRI, the acquisition of slice-level measurements over a significant time window, as a feature with the acquisition order providing a time axis for free, allowing transient events to be characterized without extra measurements.

\paragraph{Approach and contributions}
We turn the multi-view inconsistencies that conventional reconstruction discards into an explicit, interpretable signal.
\begin{itemize}[leftmargin=*,nosep]
\item \textbf{Forward model.} Per subject, every voxel of every stack is explained through its exact acquisition geometry (scanner affine transform and slice profile).
\item \textbf{Decomposition.} The model comprises a static anatomy $A$ shared by all stacks, rigid and smooth non-rigid motion, an intensity model per stack, and a sparse, signed transient field $T_v$ for changes seen in one stack only.
\item \textbf{Attribution instead of rejection.} Such changes are attributed to their acquisition time instead of averaged or rejected. Without event labels, the fit yields a transient-free reconstruction, maps of where and when each stack changed, and per-slice transient strengths over time.
\item \textbf{Transfer.} Developed on fetal MRI with inter-slice motion and uterine MRI with peristalsis, the configuration transfers unchanged to multi-site prostate MRI.
\end{itemize}

\subsection{State of the art}\label{sec:SoTA}
Existing work treats static reconstruction and the detection of inconsistencies as separate problems. Multi-view and SVR methods remove inconsistencies, quality-control methods detect them, and dedicated dynamic acquisitions measure physiology.\\
\textbf{Multi-view reconstruction.} NeRF in the Wild~\cite{Martin-Brualla_2021_CVPR} separates a static scene from a per-image transient component by discarding a learned uncertainty at rendering. RobustNeRF~\cite{Sabour_2023_CVPR} treats distractors as outliers of a robust loss, and SpotLessSplats~\cite{sabour2025spotlesssplats}, WildGaussians~\cite{kulhanek2024wildgaussians} and RobustSplat~\cite{fu2025robustsplat} carry this to 3D Gaussian splatting, with masks or uncertainties derived from residuals or pretrained features. In HybridGS~\cite{lin2024hybridgs} per-image 2D Gaussians model the transients. Transients are considered nuisances, while our goal are time-resolved transient fields as meaningful outputs.\\
\textbf{Nuisance as signal.} Computer vision has often turned artifacts into measurements, recovering depth from defocus~\cite{pentland1987new}, motion from blur~\cite{dai2008motion}, shape from shadows~\cite{savarese20073d} and hidden scenes from shadows around corners~\cite{bouman2017turning}. Closest to our setting, the line-by-line readout of a rolling-shutter camera distorts moving scenes but also encodes their motion, since every image row has its own capture time~\cite{ait2006simultaneous}. Multi-stack MRI has the same structure at a coarser scale, as every slice is acquired at a known time.\\
\textbf{Dynamic and deformable scenes.} D-NeRF~\cite{pumarola2021d} represents a moving scene as a canonical configuration and a deformation per instant, Nerfies~\cite{park2021nerfies} with an elastic regularizer. HyperNeRF~\cite{park2021hypernerf} lifts it to higher dimensions for topological changes. All rely on dense frames that each see the whole scene. In multi-stack MRI every slice has its own time, so we keep one smooth deformation per stack and assign what it cannot explain to a sparse transient field.\\
\textbf{SVR, motion and quality control.} SVR fuses stacks of thick 2D slices by alternating slice-to-volume registration with reconstruction. SVR~\cite{kuklisova2012reconstruction} introduced intensity matching and a robust statistical model to remove outliers. NiftyMIC~\cite{ebner2020automated} embeds outlier-robust super-resolution in an automated pipeline with brain localization and segmentation. SVRTK~\cite{uus2020deformable} adds deformable registration, corrects each slice by a scale and bias and weights residuals by voxel- and slice-level inlier probabilities. NeSVoR~\cite{xu2023nesvor} represents the volume as an implicit neural field and down-weights inconsistent data through pixel- and slice-level variances, and SVoRT~\cite{xu2022svort} initializes it with transformer-predicted poses. tripleSR~\cite{schlereth2025faster} fuses pre-aligned orthogonal views of the adult brain with a point-wise coordinate loss, which our forward model extends by the slice profile, motion and transients. Corrupted data are flagged by learned quality control of stacks~\cite{sanchez2024fetmrqc,xu2020semi} and volumes~\cite{sanchez2025automatic}, or detected to trigger re-acquisition~\cite{gagoski2022automated,verdera2025heron}. These approaches need labels or training data. Ours flags corrupted slices as a by-product of the reconstruction. Dedicated dynamic acquisitions quantify uterine contractions~\cite{hutter2022dynamics,verdera2025comprehensive,turk2020placental}, adding scan time, while in anatomical stacks the same events are treated as artifacts.\\
\section{Methods}
\label{sec:methods}

\subsection{Datasets}\label{sec:data}
Three different challenges serve as examples, a female pelvic and a fetal brain MRI dataset, acquired as part of ethically approved studies, and a public prostate dataset.

\noindent \textbf{Pelvis.} 123 pelvic MRI Turbo-Spin Echo examinations on 0.55T focusing on uterine abnormalities were acquired (age 18--49 years, 86 healthy, 37 pathological) with sagittal, coronal and axial stacks (0.94\,mm in-plane, 0.85\,s/slice, 4.0\,min median/stack). \textbf{Fetal brain.} 289 fetal MRI examinations (26.3--40.9 weeks gestation, 18 with pathologies), performed to assess brain development, were acquired in two interleaved passes (even followed by odd slices) with $1.48\times1.48\times4.5$\,mm resolution (1.6\,s/slice, 2.9\,min median/stack). A second cohort of 165 healthy fetuses, acquired like the fetuses with pathology, serves for the gestational-age analysis (Sec.~\ref{sec:qc}, 107 with known age, 17.1--39.3 weeks) and as a matched healthy reference for anomaly detection (Appendix~\ref{sec:anomalyexp}). \textbf{Prostate.} 300 T2-weighted prostate examinations (age 43--89 years, median 66) from the public PI-CAI dataset (fold 0)~\cite{SahaBosmaTwilt2024}, acquired at three centers on seven Siemens and Philips 1.5 and 3\,T MRIs, with $0.23--0.63$~\,mm in-plane and $3.0--3.6\,mm$ slice resolution. 87 examinations contain clinically significant cancer (47 with expert masks), 213 are considered healthy (128/85 for training/testing).

Our reconstruction is fitted to each examination alone without requiring training data. Only the anomaly detector (Appendix~\ref{sec:anomalyexp}) and the choice of design are based on several subjects, so the healthy examinations of all three datasets are split into training and test sets by a seeded rule fixed before any fit, and the test sets serve for evaluation only.

\begin{figure}[t]
\centering
\includegraphics[width=0.9\columnwidth]{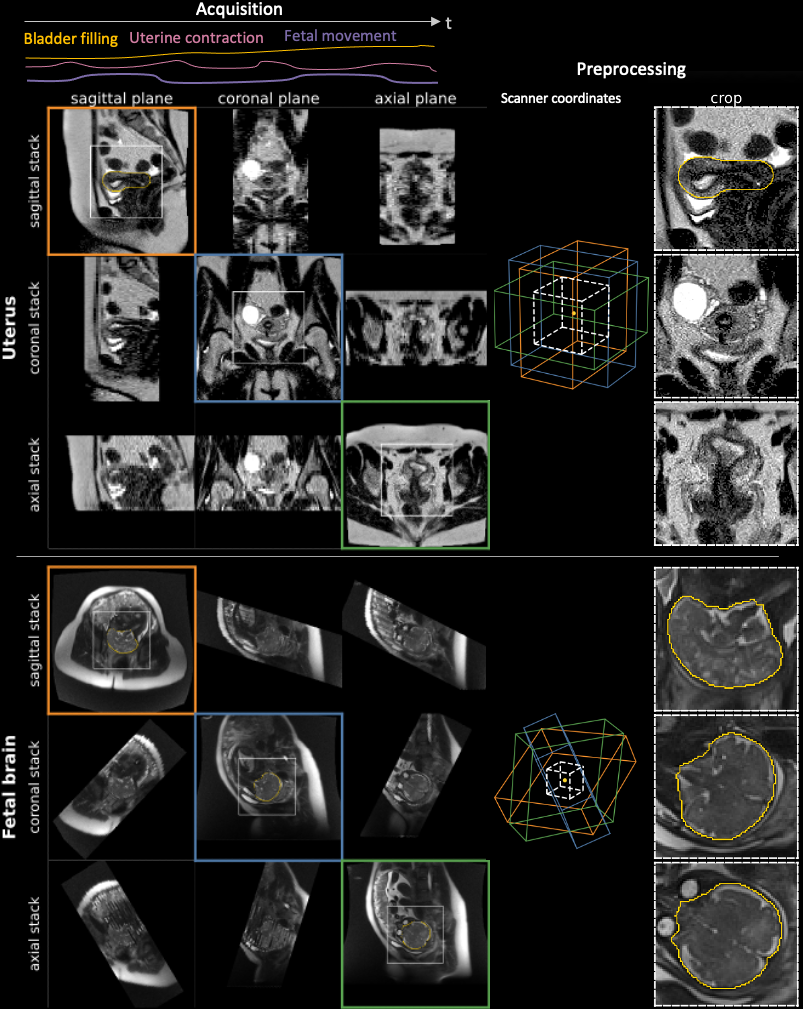}
\caption{Multi-stack acquisition and preprocessing, uterus (top) and fetal brain (bottom). Left: one stack per row, acquired consecutively while the bladder fills and the fetus moves, in its own plane on the diagonal (organ mask yellow, crop window white) and reformatted off it, where its 4.5\,mm slices appear as steps. Middle: fields of view (colored) and crop cube (white) in scanner coordinates. Right: crops cut without resampling.}
\label{fig:preprocessing}
\end{figure}

\emph{Preprocessing.} Each examination is cropped, around the uterus/gland mask of the sagittal stack with a 20\,mm margin (uterus/prostate) or around the union of the brain masks of all stacks with a 10\,mm margin (fetal brain), without resampling, by shifting its voxel-to-scanner map $M_v$. Parts of the cube outside a stack's field of view are marked and ignored. The fit is based on the whole crop (Fig.~\ref{fig:preprocessing}). 

\subsection{Problem Formulation}
\label{sec:problemform} 
\begin{figure*}[t]
\centering
\includegraphics[width=0.85\textwidth]{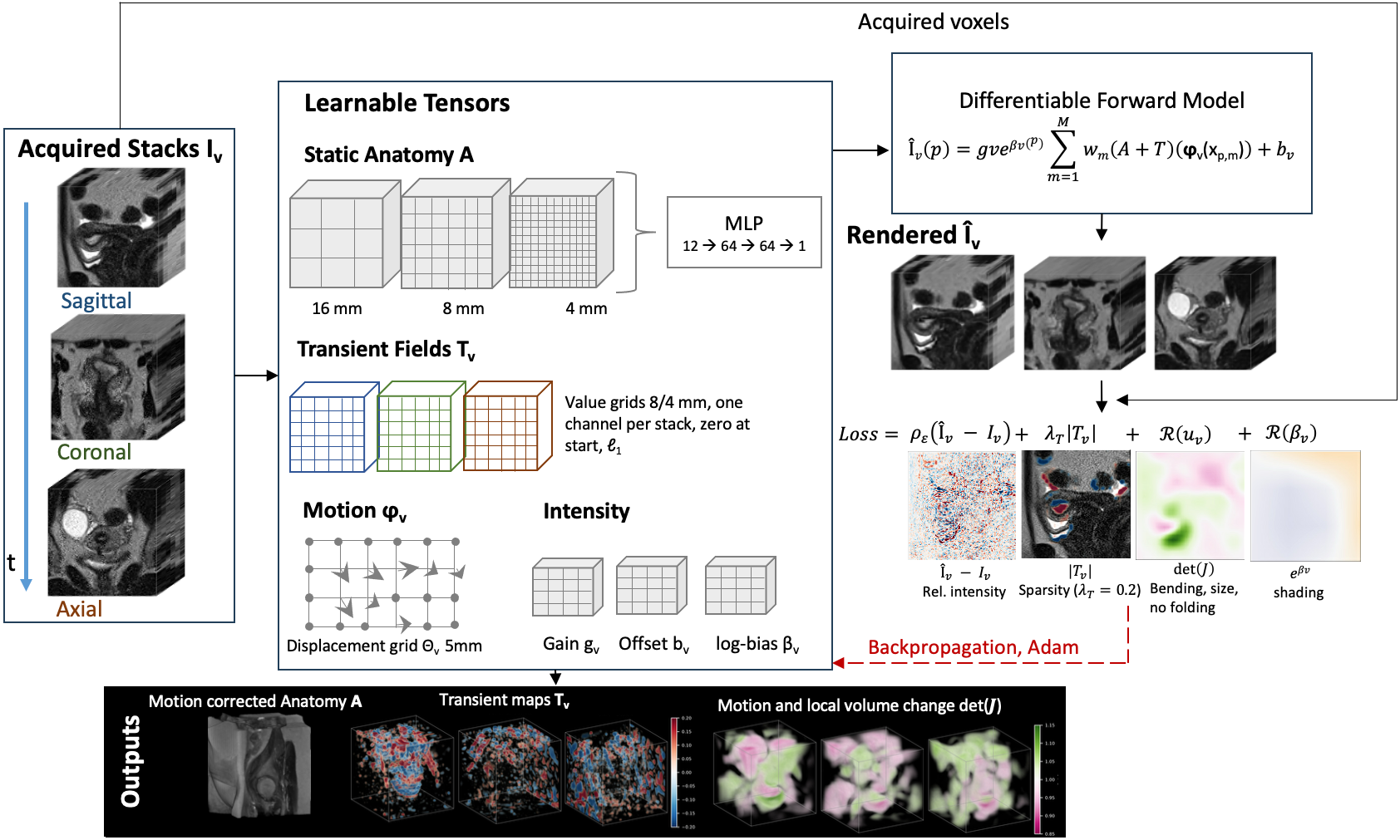}
\caption{Pipeline of our proposed per-patient decomposition. The acquired stacks are explained jointly by a static shared anatomy $A$, the rigid and non-rigid motion $\varphi_v$ of every stack, an intensity model per stack, and a sparse transient field $T_v$ for individual changes. A forward model renders every stack through its acquisition geometry, all components are fitted together to the acquired voxels in three stages.}
\label{fig:pipeline}
\end{figure*}

Each acquisition consists of $N=3$ stacks of 2D slices, acquired consecutively in different orientations, all while the imaged organ moves and changes. We fit one model per subject that predicts every voxel of every stack and separates shared anatomy and transient changes (Fig.~\ref{fig:pipeline}). 

Each stack $v \in \{1,\dots,N\}$ provides intensities $I_v(p)$ on its own voxel grid $\Omega_v$, with a known voxel-to-world affine $M_v$, slice thickness $\tau_v$ and acquisition time $t_{v,k}$ of every slice $k$. All stacks observe one static anatomy $A$, which each acquisition sees displaced by its own motion and altered only by what changed during that acquisition. Stack $v$ is then modeled as
\begin{equation}
I_v \;=\; g_v\, e^{\beta_v} \odot \mathcal{P}_v\big[(A + T_v) \circ \varphi_v\big] \;+\; b_v \;+\; n_v ,
\label{eq:generative}
\end{equation}
where all fields are defined in scanner coordinates. The static anatomy $A:\mathbb{R}^3 \to \mathbb{R}$ assigns an intensity to every point of this frame, and the transient field $T_v:\mathbb{R}^3 \to \mathbb{R}$ adds the intensity change that only stack $v$ observed. The motion $\varphi_v$ maps a point in the coordinates of stack $v$ to the point of the anatomy that the stack observed there. 
$(g_v, b_v, \beta_v)$ are gain, offset and smooth log-bias, $\mathcal{P}_v$ is the known sampling operator of the stack (its affine and slice profile, Sec.~\ref{sec:geometry}) and $n_v$ is noise. With a slice profile of zero width, no motion and no transients, Eq.~\eqref{eq:generative} reduces to the point-wise coordinate loss of tripleSR~\cite{schlereth2025faster}. Given only the subject's stacks, the task is to estimate
$\Psi = \big(A, \{\varphi_v, T_v, g_v, b_v, \beta_v\}_{v=1}^N\big)$, without training data or event labels.
\begin{equation}
\begin{aligned}
\hat\Psi \;=\; & \arg\min_{\Psi} \sum_{v} \mathcal{D}\big(I_v,\, \hat I_v(\Psi)\big) + \mathcal{R}_A(A) \\
& \quad + \sum_v \Big(\mathcal{R}_\varphi(\varphi_v) + \mathcal{R}_T(T_v) + \mathcal{R}_\beta(\beta_v)\Big),
\end{aligned}
\label{eq:estimate}
\end{equation}
with a data term $\mathcal{D}$ and regularizers $\mathcal{R}$ (see Secs.~\ref{sec:geometry} to~\ref{sec:optim}). 

\subsection{Forward Model}\label{sec:geometry}
Every stack is given in its native geometry by its voxel grid $\Omega_v$ with indices $p=(i,j,k)$, where $k$ is the slice index, its affine map $M_v$, slice spacing $\Delta_v$ and thickness $\tau_v$. All fields are defined on the scanner axes, with the origin in the organ center, on an axis-aligned box enclosing the field of view of every stack. The operator $\mathcal{P}_v$ of Eq.~\eqref{eq:generative} integrates a field $f$ along the slice normal, modeling the averaging of tissue over its thickness
\begin{equation}
\mathcal{P}_v[f](p) \;=\; \sum_{m=1}^{9} w_m\, f\big(M_v(p + \delta_m\, e_3)\big),
\label{eq:forward}
\end{equation}
where $e_3$ is the slice axis in voxel indices, the offsets $\delta_m$ are spaced evenly over $\pm 1.5$ full widths at half maximum (FWHM $= \tau_v/\Delta_v$ in slice indices, one slice thickness), and the weights $w_m$ form a normalized Gaussian~\cite{kuklisova2012reconstruction}. As $f = (A + T_v)\circ\varphi_v$, every profile point is moved by the motion of its stack before the fields are sampled. Each stack blurs only along its own slice normal.\\
\noindent \emph{Static anatomy.} The function $A$ of Eq.~\eqref{eq:generative} is evaluated at arbitrary points $x$ (profile points Eq.~\eqref{eq:forward}). It is stored as feature grids $G_l$ with four learned features per node, spaced $s_l \in \{16, 8, 4\}$\,mm, whose trilinear interpolations at $x$ ($3 \times 4=12$ features) a small MLP $f_\theta$ decodes to the intensity at $x$~\cite{xu2023nesvor} $A(x) = \operatorname{softplus}\big(f_\theta([\alpha_1 G_1(x), \dots, \alpha_L G_L(x)])\big)$.
Features and MLP weights are optimized jointly. The level weights $\alpha_l \in \{0, 1\}$ switch the levels from coarse to fine, estimating motion first on a coarse anatomy (Sec.~\ref{sec:optim}). The finest level sets the scale of detail, while the decoder can place edges within a cell. $\mathcal{R}_A = \lambda_{TV}\,\mathrm{TV}(G_L)$ is a total variation of the finest feature grid. Unlike $T_v$, $A$ is decoded by an MLP, since it is dense.\\
\noindent \emph{Motion.} $\varphi_v$ is a rigid transform followed by a smooth displacement, with rotation $\omega_v$ and translation $t_v$ for the whole stack, and a displacement $u_v$ for the deformation of the organ relative to $A$ (\eg after bladder filling).
\begin{equation}
\thickmuskip=2mu \medmuskip=2mu
\varphi_v = (\mathrm{Id} + u_v) \circ R_v,\;R_v(x) = \mathrm{Rot}(\omega_v)\,x + t_v,\;u_v = K_\sigma * \Theta_v
\label{eq:motion}
\end{equation}
$u_v$ is stored as displacement vectors $\Theta_v$ (in mm) on a 5\,mm grid, smoothed by a Gaussian $K_\sigma$ of 1.5 cells ($\approx$7.5\,mm) and interpolated trilinearly, so neighboring tissue moves together. Intra-stack motion, such as fetal movement, is not modeled by $\varphi_v$. It appears in $T_v$ as the intensity change it causes in affected slices, not as a displacement, which identifies corrupted slices (Sec.~\ref{sec:qc}). The motion parameters receive their gradients through the coordinates of Eq.~\eqref{eq:forward}, registration to the anatomy and reconstruction of the anatomy are one joint problem. The displacement is regularized by $\mathcal{R}_\varphi(\varphi_v) = \lambda_b\,\mathcal{B}(u_v) + \lambda_m \lVert u_v\rVert^2 + \lambda_f\,\mathcal{F}(u_v)$ ($\lambda_b = 10$, $\lambda_m = 10^{-4}$, $\lambda_f = 10$), which combines the bending energy $\mathcal{B}$ (squared second differences) for smoothness, a small magnitude term and a folding penalty, the mean over the grid nodes of $\mathcal{F}(u_v) = \max\big(0,\; 0.1 - \det(\mathrm{Id} + \nabla u_v)\big)$, so tissue may be compressed but not folded.\\
\noindent \emph{No reference stack.} The frame of $A$ is defined by the stacks themselves. Rotations, translations, displacements, log-gains and offsets are centered over the fitted stacks, $\omega_v \leftarrow \omega_v - \tfrac1N\sum_w \omega_w$ (same for $t_v$, $\Theta_v$, $\log g_v$, $b_v$). $A$ lies at the average pose, shape and intensity, with rotations averaged in axis-angle form and stack motion relative to it.\\
\noindent \emph{Intensity.} Before fitting, the intensities of each stack are divided by their 99th percentile. Brightness differences between the stacks are then modeled per stack by a scalar gain $g_v = e^{\gamma_v}$ and offset $b_v$ (Eq.~\eqref{eq:generative}) and a smooth log-bias $\beta_v$, stored on a $4\times4\times3$ grid in the stack's voxel index space, interpolated trilinearly and regularized by $\mathcal{R}_\beta(\beta_v) = \lambda_\beta \lVert \beta_v \rVert^2$ ($\lambda_\beta = 0.05$). These differences are thus explained neither by $A$ nor by $T_v$.

\subsection{Transient Fields}\label{sec:inconsistency}
Once motion and intensity are accounted for, the remaining differences between the stacks are changes of single acquisitions, such as contractions or motion corruption. Instead of down-weighting them, Eq.~\eqref{eq:generative} models them by a signed field per stack, $T_v(x) = H_1^{(v)}(x) + H_2^{(v)}(x)$, with values on regular grids of 8 and 4\,mm for extended changes and local detail, interpolated trilinearly. The grids are the same for all datasets. Their finest spacing is close to one slice (3.0--4.5\,mm), so a change confined to one slice reaches at most its spatial neighbors, which in an interleaved acquisition were acquired at another time. Unlike $A$, $T_v$ has no decoder, so every value affects only its own cells and $\mathcal{R}_T(T_v) = \lambda_T\,\mathbb{E}_x|T_v(x)| + \lambda_{\mathrm{TV},T}\,\mathrm{TV}(H_2^{(v)})$ acts locally. The $\ell_1$ term ($\lambda_T = 0.2$) keeps $T_v$ zero wherever no transient is required, and the total variation ($\lambda_{\mathrm{TV},T} = 10^{-3}$) suppresses isolated values. As $T_v$ lives in the frame of $A$ and is added before motion and slice sampling, a transient has a position in the anatomy and is observed by stack $v$ through its slice profile and motion.

\emph{Robust reweighting.} Where stack $v$ needs a transient, $A$ should follow the other stacks. Instead of down-weighting the data, as robust losses and outlier rejection in SVR do, we route the gradient. We replace $A$ by
\begin{equation}
\tilde A_v = w_v A + (1-w_v)\,\mathrm{sg}[A], \qquad w_v = \frac{1}{1 + (T_v/c)^2},
\label{eq:robust}
\end{equation}
with the stop-gradient $\mathrm{sg}[\cdot]$ and $c = 0.05$. The prediction is unchanged, but the gradients through $A$ are scaled by this Cauchy weight, so a single-stack change that $A$ absorbed before $T_v$ was fitted is pushed back into $T_v$. Unlike outlier rejection in SVR, the disagreement is kept.

\subsection{Objective and Optimization}\label{sec:optim}
We minimize Eq.~\eqref{eq:estimate} with the described regularizers and the mean Charbonnier penalty of the residual $r_v(p) = \hat I_v(p) - I_v(p)$ as data term,
\begin{equation}
\mathcal{D}\big(I_v, \hat I_v\big) \;=\; \tfrac{1}{N}\,\mathbb{E}_{p \in \Omega_v}\,\sqrt{r_v(p)^2 + \epsilon^2}.
\label{eq:loss}
\end{equation}
In the first two stages, $\epsilon = 10^{-3}$ makes $\mathcal{D}$ close to an $\ell_1$ loss, so changes that no $T_v$ explains yet pull $A$ and motion less than a quadratic loss. With $T_v$, $\epsilon=0.05$ ($~$ noise level), so $T_v$ follows the local mean of the residual and $\lambda_T$ acts as a threshold. Voxels stored as exactly zero (clipping at export) only penalize predictions above the clip level. Weights: $\lambda_T = 0.2$, $c = 0.05$, $\lambda_{TV} = \lambda_{\mathrm{TV},T} = 10^{-3}$, $\lambda_b = 10$, $\lambda_m = 10^{-4}$, $\lambda_f = 10$, $\lambda_\beta = 0.05$. Each stage (1000 iterations) optimizes its parameters jointly with Adam and a decaying cosine schedule, on 16{,}384 random voxels per stack and iteration. The order of the stages decides which term may explain what (Appendix~\ref{sec:ablation}). (i) \emph{Rigid}: $A$, $R_v$ and the intensities (learning rate $10^{-2}$), with the levels of $A$ switched on at 15\% intervals. (ii) \emph{Non-rigid}: $u_v$ is added ($5\cdot10^{-3}$, $10\times$ for $\Theta_v$) and $A$ restarts coarse to fine, reaching its finest level at 60\%, so that the deformation is estimated before $A$ can store a compromise between the stacks. (iii) \emph{Transients}: $T_v$ is added with robust reweighting ($10\times$ the rate of $A$) while everything else keeps adapting, so $T_v$ receives only what anatomy and motion cannot explain.

\subsection{Quantifying dynamic activity}\label{sec:dynamics}
The known acquisition time $t_{v,k}$ of each slice allows spatio-temporal description. We derive one measure per slice and three spatial maps on the 1\,mm grid of $A$, evaluated only within tissue covered by $\geq2$ stacks, since motion and transients are defined by inter-stack differences. \emph{Displacement.} $\varphi_v^{-1}(y) - y$ (in mm) is where the tissue at $y$ was during acquisition $v$ relative to $A$, split into the rigid displacement and the non-rigid remainder. \emph{Local volume.} $V_v = 1/\det(\mathrm{Id} + \nabla u_v)$, with the Jacobian from finite differences on the displacement grid, is the volume of the tissue in acquisition $v$ relative to $A$. \emph{Transient map.} $T_v$, in normalized intensity, presents where acquisition $v$ saw a change that the others did not, brighter where $T_v(x)>0$ and darker where $T_v(x)<0$. \emph{Transient strength per slice.} Rendering stack $v$ through Eq.~\eqref{eq:generative} with the same motion and intensities, once with and without $T_v$, gives the transient as acquired, $\Delta_v = \hat I_v^{\,A+T_v} - \hat I_v^{\,A}$, and its strength in slice $k$, $s_{v,k} \;=\; \mathbb{E}_{p \in \Omega_{v,k}}\big|\Delta_v(p)\big| \,\big/\, \mathbb{E}_{p \in \Omega_{v,k}}\, I_v(p)$, with $\Omega_{v,k}$ the organ mask in slice $k$, dilated by 10\,mm. A slice is flagged when $s_{v,k}$ exceeds the 95th percentile over all slices outside this evaluation.

\section{Experiments}\label{sec:experiments}
In-vivo, neither the anatomy nor the transients are known, so no single experiment can validate the decomposition. We therefore test each claim with bespoke experiments (Appendix~\ref{app:overview}). Whether $A$ is a faithful anatomy is tested by comparing predictions with held-out stacks, the analogue of novel-view synthesis. Whether changes are attributed to the right component is tested on a digital phantom with known anatomy, motion and transients (Sec.~\ref{sec:phantom}). Whether $T_v$ captures what changed during one acquisition is tested against synthetically corrupted slices and expert ratings (Sec.~\ref{sec:qc}).  Whether $A$ serves as input for screening pathology is tested on synthetic and real lesions (Appendix~\ref{sec:anomalyexp}), and the ablation of each model component is given in Appendix~\ref{sec:ablation}. Every setting was fixed on the uterine and fetal data before evaluation, and the prostate data test generalization to another organ, site and scanner.

\subsection{Evaluation protocols}\label{sec:protocols}
One configuration is employed for all cases (Sec.~\ref{sec:optim}).\\
\noindent \emph{Held-out stack.} One stack $h$ is left out while $A$, the motion of the other stacks and their transients are fitted as above. Then $A$ is frozen and only the rigid and non-rigid motion and the intensity model of stack $h$ are fitted (400 iterations, no transient), and $h$ is predicted through Eq.~\eqref{eq:generative}. The baseline is the mean of the motion-corrected other stacks at the same points, without a slice profile. A stack is registered anew to the volume (rigid, non-rigid, and gain, offset and smooth bias over the whole stack), simulated through its slice profile and compared by NCC, SSIM and PSNR in a common region for every stack the volume was reconstructed from (seen stacks) and for the held-out stack. SVRTK~\cite{uus2020deformable} and NeSVoR~\cite{xu2023nesvor} receive exactly the same stacks and our crops and masks. SVRTK reconstructs in the space of a template stack, chosen via the largest brain mask (fetal) or sagittal stack (uterus and prostate). NeSVoR uses SVoRT registration~\cite{xu2022svort} for the fetal brain and stack registration for the uterus and prostate. Fetal volumes are compared within the brain mask and uterine/prostate volumes within the uterus/prostate +10\,mm margin. Appendix~\ref{app:heldout} presents an example for the fetal brain and the uterus. A lesion insertion benchmark for the inputs of the anomaly detector is described in Appendix~\ref{sec:anomalyexp}. 

\noindent \emph{Metrics and statistics.} Reconstructions are compared by NCC and SSIM and PSNR (dB). Slice flagging is scored by AUROC with 95\% bootstrap confidence intervals over stacks. To compare at matched resolution, prediction and held-out stack are also low-pass filtered in-plane by a common Gaussian ($\sigma = 1$, 2 and 4\,mm, normalized within the scored region), and the differences between successive filter levels give frequency bands (0--1, 1--2, 2--4 and above 4\,mm), each scored by NCC. Methods are compared by the Wilcoxon signed-rank test, a method is better only if the median per-patient difference favors it and $p < 0.05$.

\subsection{Separating transients}\label{sec:phantom}
\begin{figure}[t]
\centering
\includegraphics[width=\columnwidth]{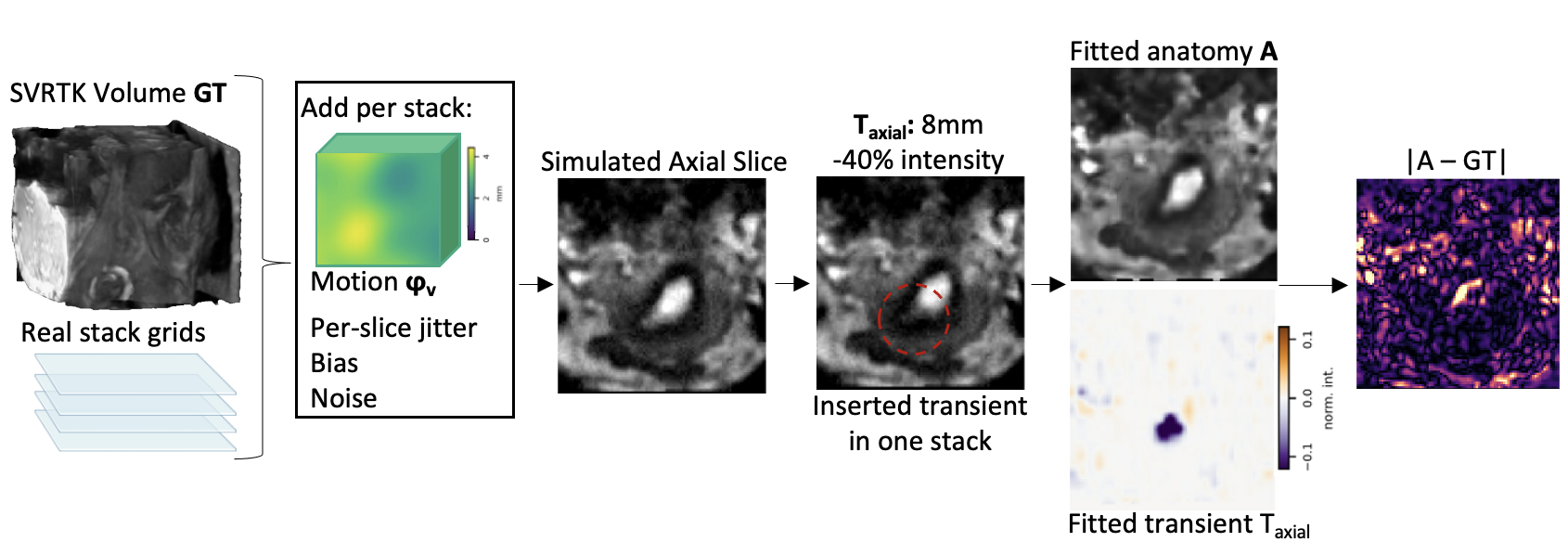}
\caption{Digital phantom (uterus). Left to right: known input, ground-truth anatomy $\mathrm{GT}$ (an SVRTK reconstruction), known motion of one stack, that stack as simulated and with the inserted changes, the fit, its anatomy $A$, transient field $T_v$ and $|A - \mathrm{GT}|$, with the share of the transient found in $T_v$ and leaked into $A$.}
\label{fig:phantom}
\end{figure}

\begin{figure}[!h]
\centering
\includegraphics[width=0.80\columnwidth]{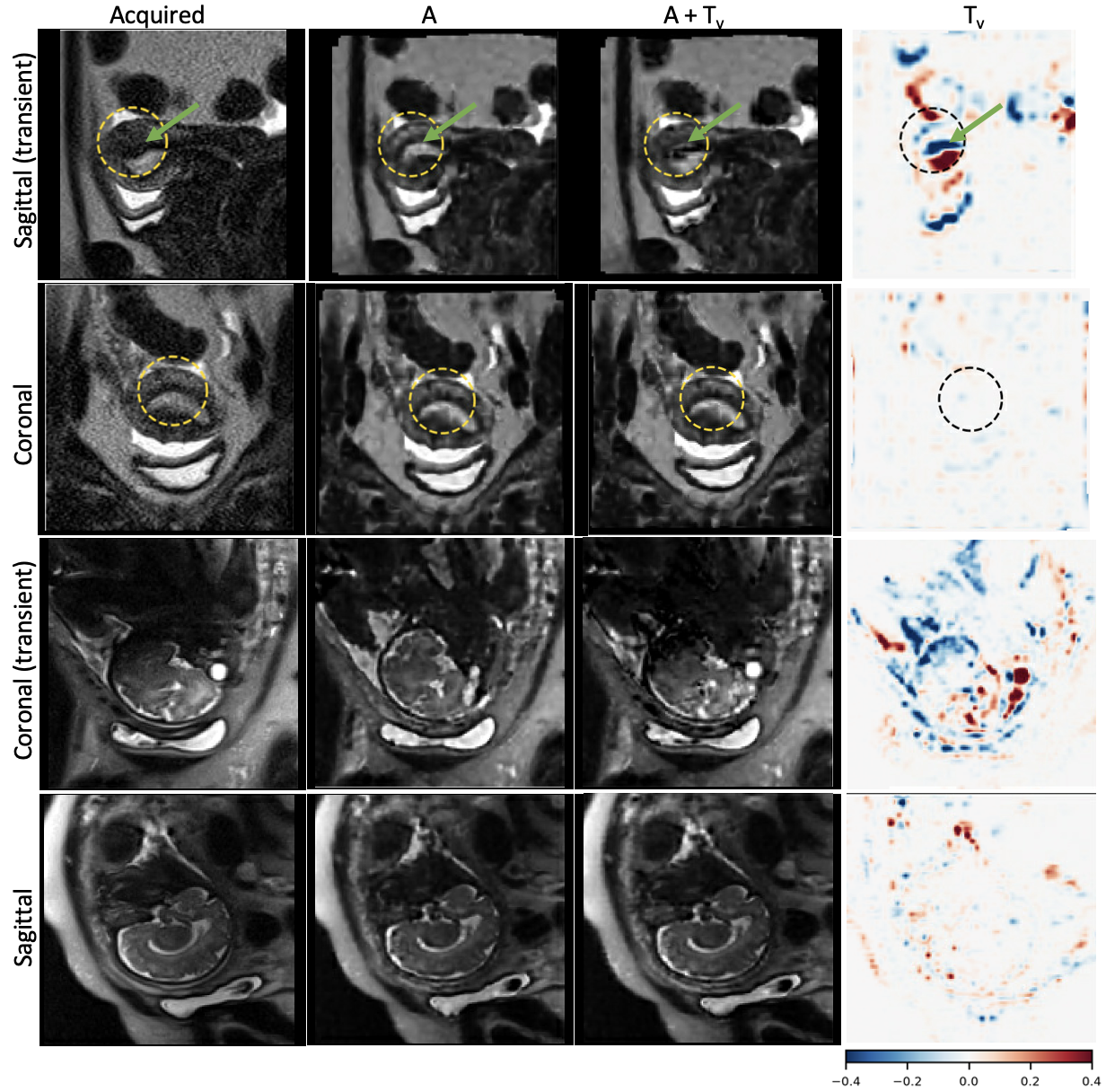}
\caption{Changes seen by one stack, in a uterine (top) and a fetal exam (bottom). Columns: acquired slice, anatomy $A$, $A + T_v$ and the transient field $T_v$ of that stack. Uterus: a dark band (arrows) appears only in the sagittal stack and is recovered in $T_v$. Fetal brain: coronal signal loss is isolated in $T_v$ and excluded from $A$.}
\label{fig:contraction}
\end{figure}

\begin{table}[t]
\centering\small
\caption{Digital phantom (per dataset 20 phantoms built from 20 patients, mean $\pm$ SD, transients as median). Attribution is the projection of the fitted change onto the inserted one within its region (1 = fully recovered, $>1$ = fitted change exceeds the inserted one).}
\label{tab:phantom}
\resizebox{\columnwidth}{!}{
\begin{tabular}{lcc}
\toprule
 & Uterus & Prostate \\
\midrule
$A$ vs.\ ground truth, NCC $\uparrow$ & $0.981 \pm 0.007$ & $0.938 \pm 0.015$ \\
Motion error, mm (applied $3.3 / 3.6$) $\downarrow$ & $0.27 \pm 0.06$ & $0.24 \pm 0.06$ \\
Static change found in $A$ (ideal 1) & $1.00 \pm 0.07$ & $1.08 \pm 0.06$ \\
Transient in own $T_v$, $r = 5/8/12$\,mm $\uparrow$ & $0.58/0.64/0.68$ & $0.77/0.77/0.76$ \\
Transient leaked into $A$ $\downarrow$ & $0.06$--$0.13$ & $0.04$--$0.06$ \\
\bottomrule
\end{tabular}
}
\end{table}
The SVRTK reconstructions of 20 patients per dataset serve as synthetic ground-truth anatomies $\mathrm{GT}$ to allow simulation of known transients. The simulation differs from our forward model, with a trapezoidal slice profile, in-plane blur, per-slice jitter of 0.3\,mm, a polynomial bias of $\pm10\%$ and Rician noise. Each patient's three stacks are simulated under known, centered motion ($\pm3^\circ$, $\pm4$\,mm, smooth deformation of 1--3\,mm), once without changes and twice with changes: two intensity transients (spheres of 5, 8 or 12\,mm, contrast $\pm40\%$), each in one stack, a local compression of up to 3\,mm in one stack that mimics a contraction, and a static change in all three stacks. All cases share motion and noise, so every change is isolated by subtracting the fit of the unchanged case. On both organs, $A$ matches $\mathrm{GT}$ closely and the motion is recovered to 0.27 and 0.24\,mm (Table~\ref{tab:phantom}). Static changes stay in $A$, while only a small share of the transients leaks into $A$. On the uterus the transient field captures larger transients more completely. On the prostate, with thinner slices of 3.0--3.6\,mm instead of 4.5\,mm, 0.77 of a transient is found in its field at every radius. In patients, the strongest transients confined to the uterine wall and seen by a single stack resemble contractions, as the dark band in the sagittal stack only illustrates (Fig.~\ref{fig:contraction}). They are captured only by the corresponding transient field. Averaged into the anatomy, such a band could resemble a lesion.

\subsection{Reconstruction}\label{sec:recon}
\emph{Held-out prediction against the mean of the stacks.} On the uterus (123 patients, 369 cases), $A$ predicts the held-out stack with median NCC 0.897, against 0.870 for the motion-corrected mean of the other stacks and 0.835 without motion correction, and is better than the motion-corrected mean in 119/123 patients (median per-patient difference $+0.023$, $p < 10^{-20}$). On the fetal brain (83 fetuses, 249 cases), $A$ reaches 0.771 against 0.746 and 0.621, better than the motion-corrected mean in 66/83 fetuses ($+0.024$, $p < 10^{-7}$). On the prostate (85 patients, 255 cases), the result is the same, with 0.847 against 0.828 and 0.727. Motion correction and the fusion into one anatomy thus both contribute to the prediction.

\emph{Consistency with seen and held-out stacks.} SVRTK and NeSVoR reproduce the seen uterine stacks more closely (median NCC 0.951 and 0.946 against 0.934, SSIM 0.791 and 0.757 against 0.608, Table~\ref{tab:recon}), because they average everything the stacks contain into one volume, and our fit without transient fields also reaches 0.948. They also reconstruct at 0.8\,mm and keep the in-plane detail of the 0.94\,mm slices, while the finest grid of $A$ is 4\,mm, which favors them on seen stacks. On the fetal brain, $A$ is more consistent with the seen stacks than SVRTK and NeSVoR (median NCC 0.893 against 0.868 and 0.835, better in 78 and 81 of 83 fetuses), and restricting our fit to the brain masks lowers this to 0.860, so modeling the whole field of view also improves the brain itself. On the prostate, $A$ is more consistent in NCC (0.909 against 0.900 and 0.834) and PSNR, while SVRTK again reaches a higher SSIM (0.577 against 0.527), in line with its finer grid.
The order reverses for the held-out stacks. On the uterus, SVRTK and NeSVoR reach 0.874 and 0.863, below $A$ in 123 and 122 of 123 patients ($p < 10^{-20}$, similar for SSIM and PSNR), and barely improve on the mean of the stacks, so the changes they average in fit the seen stacks but not a new acquisition. This is not due to the coarser grid of $A$, which also stays ahead after a common in-plane low-pass (2\,mm: 118 and 121 of 123 patients) and in the finest frequency band (0--1\,mm: 0.31 against 0.24 and 0.26). On the prostate, on which no setting was chosen, SVRTK and NeSVoR reach 0.795 and 0.757, below $A$ in 85 and 84 of 85 patients ($p < 10^{-14}$), and in all 85 after the low-pass. On the fetal brain, $A$ predicts as well as SVRTK (0.763, $p = 0.37$) and better than NeSVoR (0.747, $p < 10^{-4}$), with NeSVoR agreeing better in the bands below 4\,mm. Since the evaluation fits gain, offset and bias over the whole stack, it penalizes volumes that are empty outside the brain. With this fit restricted to the brain (a sensitivity analysis declared after the primary result), $A$ is better than both (0.804 against 0.785 and 0.773, $p < 10^{-3}$) and NeSVoR's advantage in the fine bands vanishes. Fitting our model only inside the masks lowers its prediction to 0.749, so the surrounding tissue helps to reconstruct the brain.

\begin{table}[t]
\centering\footnotesize
\setlength{\tabcolsep}{2.5pt}
\caption{Reconstruction from identical inputs, median over stacks. All volumes are registered to each stack and simulated through its slice profile. Mean of stacks: the mean of the other stacks at the points of the held-out stack, with (aligned) and without (unaligned) our motion correction. Uterus: 123 patients. Fetal brain: 83 fetuses. $^*$: sensitivity analysis with gain, offset and bias of the registration fitted inside the brain only. Prostate: 85 patients. Bold: best value per row.}
\label{tab:recon}
\resizebox{\columnwidth}{!}{
\begin{tabular}{@{}cllccccc@{}}
\toprule
 & & & Ours ($A$) & SVRTK & NeSVoR & \multicolumn{2}{c}{Mean of stacks} \\
\cmidrule(lr){7-8}
 & & & & & & aligned & unaligned \\
\midrule
\multirow{6}{*}{\rotatebox{90}{Uterus}}
 & Seen & NCC $\uparrow$ & 0.934 & \textbf{0.951} & 0.946 & -- & -- \\
 & & SSIM $\uparrow$ & 0.608 & \textbf{0.791} & 0.757 & -- & -- \\
 & & PSNR $\uparrow$ & 23.8 & \textbf{24.7} & 24.1 & -- & -- \\
 & Held out & NCC $\uparrow$ & \textbf{0.897} & 0.874 & 0.863 & 0.870 & 0.835 \\
 & & SSIM $\uparrow$ & \textbf{0.486} & 0.446 & 0.446 & 0.432 & 0.386 \\
 & & PSNR $\uparrow$ & \textbf{21.9} & 20.9 & 20.7 & 20.8 & 19.6 \\
\midrule
\multirow{8}{*}{\rotatebox{90}{Fetal brain}}
 & Seen & NCC $\uparrow$ & \textbf{0.893} & 0.868 & 0.835 & -- & -- \\
 & Seen$^*$ & NCC $\uparrow$ & \textbf{0.900} & 0.884 & 0.869 & -- & -- \\
 & & SSIM $\uparrow$ & 0.783 & \textbf{0.806} & 0.783 & -- & -- \\
 & & PSNR $\uparrow$ & \textbf{25.1} & 24.9 & 24.4 & -- & -- \\
 & Held out & NCC $\uparrow$ & \textbf{0.771} & 0.763 & 0.747 & 0.746 & 0.621 \\
 & Held out$^*$ & NCC $\uparrow$ & \textbf{0.804} & 0.785 & 0.773 & 0.744 & 0.614 \\
 & & SSIM $\uparrow$ & \textbf{0.616} & 0.593 & 0.588 & 0.559 & 0.450 \\
 & & PSNR $\uparrow$ & \textbf{22.8} & 22.5 & 22.2 & 21.0 & 19.8 \\
\midrule
\multirow{6}{*}{\rotatebox{90}{Prostate}}
 & Seen & NCC $\uparrow$ & \textbf{0.909} & 0.900 & 0.834 & -- & -- \\
 & & SSIM $\uparrow$ & 0.527 & \textbf{0.577} & 0.451 & -- & -- \\
 & & PSNR $\uparrow$ & \textbf{21.5} & 20.7 & 18.7 & -- & -- \\
 & Held out & NCC $\uparrow$ & \textbf{0.847} & 0.795 & 0.757 & 0.828 & 0.727 \\
 & & SSIM $\uparrow$ & \textbf{0.432} & 0.369 & 0.311 & 0.420 & 0.334 \\
 & & PSNR $\uparrow$ & \textbf{19.5} & 18.0 & 16.7 & 18.8 & 16.9 \\
\bottomrule
\end{tabular}
}
\end{table}

\emph{Practicality.} Our method ran on all three datasets with one configuration and without training data, atlas or template. NeSVoR's learned registration (SVoRT) is trained on fetal brains and had to be replaced by stack registration for the other organs, and SVRTK needs a reference stack per dataset and exceeded 32\,GB of memory on 3/123 uterine patients. A fit of ours takes 1.6--2.0\,min on one A100 GPU (peak memory below 1\,GB), against 7.3--7.9\,min for NeSVoR on the same GPU type and 1.8--6.9\,min for SVRTK on 8 CPU cores.

\subsection{Transients as a signal: slice-level quality control}\label{sec:qc} 
A corrupted slice is a transient in its simplest form, a change recorded by only one acquisition. The conventional visual inspection is slow, and excluding whole stacks discards unaffected slices. We test whether $T_v$ finds these slices without labels and whether $A$ stays usable without excluding them. \emph{Corruption benchmark.} 20 healthy fetuses were refitted after three slices per stack (180 in total) were corrupted by signal dropout, in-plane motion with ghosting or displaced tissue. The transient strength separates corrupted from clean slices with AUROC 0.90 (dropout 0.99, displaced tissue 0.92, ghosting 0.79). With only two fitted stacks, $T_v$ still reaches 0.84, and a new stack checked against the anatomy of two earlier ones 0.81, corresponding to the setting of real-time quality control. \emph{Real data.} $T_v$ raises the consistency of the model with the acquired data from NCC 0.910 to 0.963. Flagged slices cluster in acquisition time rather than in space. 40\% have a flagged neighbor in acquisition order, against 28\% in spatial order (867 interleaved stacks). Two permutation tests rearrange the flags within each stack, at random or by circular shifts in spatial order that keep their spatial clustering, and predict fewer (28\% and 37\%, both $p < 0.001$), consistent with motion events during the interleaved acquisition.

\begin{figure}[t]
\centering
\includegraphics[width=\columnwidth]{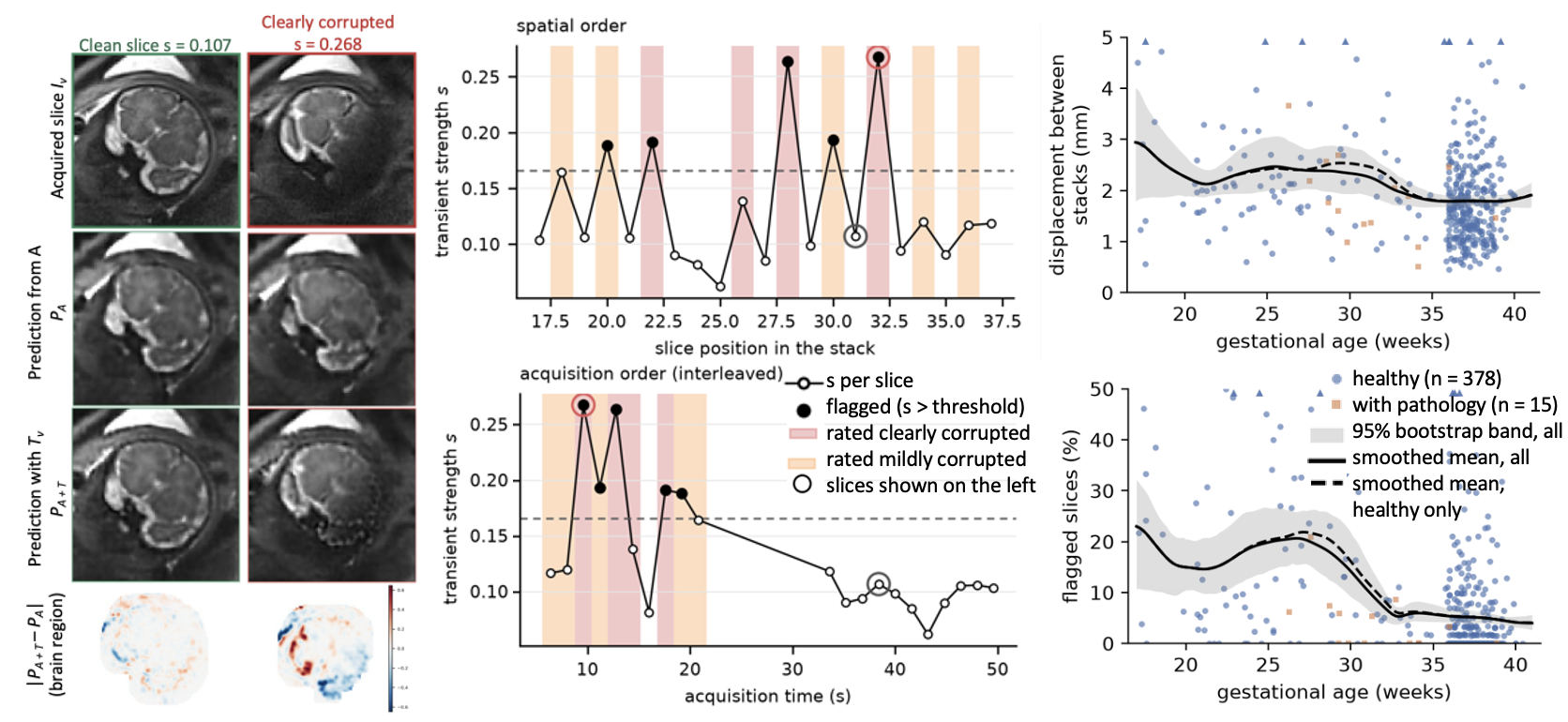}
\caption{Fetal slice-level quality control. Left: a clean and a corrupted slice of one stack ($I_v$), predicted from anatomy alone ($P_A$) and with the transient ($P_{A+T}$), and their difference. Middle: transient strength $s$ of every slice in spatial (top) and acquisition order (bottom), with expert ratings (red: clearly, orange: mildly corrupted) for comparison only. Right: brain displacement (top) and share of flagged slices (bottom) against gestational age, smoothed mean over all/healthy fetuses, points above the range as triangles.}
\label{fig:qc}
\end{figure}

\emph{Expert ratings.} Two experts marked clearly and mildly corrupted slices in 30 fetuses (90 stacks), each fetus rated by one of them. Transient strength separates the 65 clearly corrupted from 1731 clean slices with AUROC 0.94 (95\% CI 0.91--0.96), and with 0.91 including the 45 mildly corrupted slices. We flag a slice when $s$ exceeds the 95th percentile over the 239 fetal examinations outside any slice-quality evaluation. This flags 52\% (95\% CI 37--67) of the clearly corrupted slices at 3.5\% false positives, and 58\% of the synthetically corrupted slices at 8.8\%, with the 90th and 98th percentiles trading sensitivity against false positives as expected. In Fig.~\ref{fig:qc}, the flagged slices are spatially separated but close in acquisition time, as expected for an interleaved acquisition. \emph{Keeping instead of excluding.} Excluding the 26 stacks with a clearly corrupted slice would leave 13 of 30 fetuses with fewer than three stacks and discard 467 clean slices, while our model keeps every stack and flags affected slices. Slices predicted from $A$ recover the original content of corrupted slices (ghosting NCC 0.72 against 0.55, dropout PSNR 17.9 against 15.4\,dB), as does a fit without $T_v$ (0.74, 18.6\,dB), so $T_v$ serves to flag these slices, not to restore them. \emph{Fetal motion over gestation.} In the second cohort, the share of flagged slices decreases with gestational age ($\rho = -0.30$, $p = 0.002$, adjusted for protocol and slice thickness $-0.29$), from about 20\% before 29 weeks to about 5\% from 33 weeks on (Fig.~\ref{fig:qc}, right), consistent with literature~\cite{ji2023fetal}. The decrease holds in all 378 healthy fetuses adjusted for cohort ($\rho = -0.15$, $p = 0.004$). The displacement of the brain between the stacks presents no robust relation to gestational age.

\subsection{Limitations.} $T_v$ collects whatever one stack contradicts after smooth motion, intensity and the slice profile are accounted for, so it is not specific to motion, and we treat flagged slices as corrupted with motion as their likely cause. The finest grid of $A$ is 4\,mm, coarser in-plane than SVR volumes. As input to a frozen anomaly detector, $A$ does not localize prostate cancer on T2-weighted images alone (Appendix~\ref{sec:anomalyexp}).

\section{Conclusion}\label{sec:conclusion}
We treated the disagreement between stacks of a multi-stack MRI exam as a quantity to estimate rather than an outlier to remove. Fitted per subject, without training data and with one configuration, our decomposition separates a static anatomy, stack motion and transient fields. The anatomy predicts held-out acquisitions at least as well as slice-to-volume reconstruction from identical inputs, and better on uterine and prostate MRI. The transient fields flag corrupted slices and place them in acquisition time. The model transferred unchanged to prostate MRI from other sites, vendors and field strengths. What reconstruction discards thus becomes a per-acquisition record of what changed and when, a basis for checking slices during the examination, quantifying contractions and studying fetal movement, beyond a sharper volume.\\

\noindent \textbf{Acknowledgment} The authors thank all pregnant women and their families for participating in this study. The authors thank the research midwives and obstetric fellows for their invaluable efforts in recruiting and caring for the women in this study, as well as the radiographers for their participation in the acquisition of these datasets. This work was supported by the High Tech Agenda of the Free State of Bavaria, DFG Heisenberg funding [502024488], ERC Starting grant EARTHWORM [101165242], ERC PoC grant SYNCWORM [101293293], BMFTR grant LYNX [031L0329], the CAIMED consortium [ZN4257], MRC Program Grant MiBirth [MR/X010007/1], and funding from the StMGP as part of the project “EndoKI”. M.L.\ receives funding from the Berlin Institute of Health (Junior Digital Clinician Scientist Grant). The authors have no competing interests to declare.
{
    \small
    \bibliographystyle{ieeenat_fullname}
    \bibliography{main}

@String(CVPR= {IEEE Conf. Comput. Vis. Pattern Recog.})

@String(ICCV= {Int. Conf. Comput. Vis.})

@String(AAAI = {AAAI})

@String(CVPR  = {CVPR})

@String(ICCV  = {ICCV})

@InProceedings{Sabour_2023_CVPR,
    author    = {Sabour, Sara and Vora, Suhani and Duckworth, Daniel and Krasin, Ivan and Fleet, David J. and Tagliasacchi, Andrea},
    title     = {RobustNeRF: Ignoring Distractors With Robust Losses},
    booktitle = {Proceedings of the IEEE/CVF Conference on Computer Vision and Pattern Recognition (CVPR)},
    month     = {June},
    year      = {2023},
    pages     = {20626-20636}
}

@InProceedings{Martin-Brualla_2021_CVPR,
    author    = {Martin-Brualla, Ricardo and Radwan, Noha and Sajjadi, Mehdi S. M. and Barron, Jonathan T. and Dosovitskiy, Alexey and Duckworth, Daniel},
    title     = {NeRF in the Wild: Neural Radiance Fields for Unconstrained Photo Collections},
    booktitle = {Proceedings of the IEEE/CVF Conference on Computer Vision and Pattern Recognition (CVPR)},
    month     = {June},
    year      = {2021},
    pages     = {7210-7219}
}

@InProceedings{lin2024hybridgs,
    author = {Jingyu, Lin and Jiaqi, Gu and Lubin, Fan and Bojian, Wu and Yujing, Lou and Renjie, Chen and Ligang, Liu and Jieping, Ye},
    title = {HybridGS: Decoupling Transients and Statics with 2D and 3D Gaussian Splatting},
    booktitle = {CVPR},
    year = {2025}
}

@inproceedings{pumarola2021d,
  title={D-nerf: Neural radiance fields for dynamic scenes},
  author={Pumarola, Albert and Corona, Enric and Pons-Moll, Gerard and Moreno-Noguer, Francesc},
  booktitle={2021 IEEE/CVF Conference on Computer Vision and Pattern Recognition (CVPR)},
  pages={10313--10322},
  year={2021},
  organization={IEEE}
}

@inproceedings{park2021nerfies,
  title={Nerfies: Deformable neural radiance fields},
  author={Park, Keunhong and Sinha, Utkarsh and Barron, Jonathan T and Bouaziz, Sofien and Goldman, Dan B and Seitz, Steven M and Martin-Brualla, Ricardo},
  booktitle={2021 IEEE/CVF International Conference on Computer Vision (ICCV)},
  pages={5845--5854},
  year={2021},
  organization={IEEE}
}

@article{park2021hypernerf,
  title={Hypernerf: A higher-dimensional representation for topologically varying neural radiance fields},
  author={Park, Keunhong and Sinha, Utkarsh and Hedman, Peter and Barron, Jonathan T and Bouaziz, Sofien and Goldman, Dan B and Martin-Brualla, Ricardo and Seitz, Steven M},
  journal={arXiv preprint arXiv:2106.13228},
  year={2021}
}

@article{sabour2025spotlesssplats,
  title={Spotlesssplats: Ignoring distractors in 3d gaussian splatting},
  author={Sabour, Sara and Goli, Lily and Kopanas, George and Matthews, Mark and Lagun, Dmitry and Guibas, Leonidas and Jacobson, Alec and Fleet, David and Tagliasacchi, Andrea},
  journal={ACM Transactions on Graphics},
  volume={44},
  number={2},
  pages={1--11},
  year={2025},
  publisher={ACM, New York, NY}
}

@article{kulhanek2024wildgaussians,
  title={Wildgaussians: 3d gaussian splatting in the wild},
  author={Kulhanek, Jonas and Peng, Songyou and Kukelova, Zuzana and Pollefeys, Marc and Sattler, Torsten},
  journal={arXiv preprint arXiv:2407.08447},
  year={2024}
}

@inproceedings{fu2025robustsplat,
  title={Robustsplat: Decoupling densification and dynamics for transient-free 3dgs},
  author={Fu, Chuanyu and Zhang, Yuqi and Yao, Kunbin and Chen, Guanying and Xiong, Yuan and Huang, Chuan and Cui, Shuguang and Cao, Xiaochun},
  booktitle={2025 IEEE/CVF International Conference on Computer Vision (ICCV)},
  pages={27126--27136},
  year={2025},
  organization={IEEE}
}

@article{kuklisova2012reconstruction,
  title={Reconstruction of fetal brain MRI with intensity matching and complete outlier removal},
  author={Kuklisova-Murgasova, Maria and Quaghebeur, Gerardine and Rutherford, Mary A and Hajnal, Joseph V and Schnabel, Julia A},
  journal={Medical image analysis},
  volume={16},
  number={8},
  pages={1550--1564},
  year={2012},
  publisher={Elsevier}
}

@article{ebner2020automated,
  title={An automated framework for localization, segmentation and super-resolution reconstruction of fetal brain MRI},
  author={Ebner, Michael and Wang, Guotai and Li, Wenqi and Aertsen, Michael and Patel, Premal A and Aughwane, Rosalind and Melbourne, Andrew and Doel, Tom and Dymarkowski, Steven and De Coppi, Paolo and others},
  journal={NeuroImage},
  volume={206},
  pages={116324},
  year={2020},
  publisher={Elsevier}
}

@article{uus2020deformable,
  title={Deformable slice-to-volume registration for motion correction of fetal body and placenta MRI},
  author={Uus, Alena and Zhang, Tong and Jackson, Laurence H and Roberts, Thomas A and Rutherford, Mary A and Hajnal, Joseph V and Deprez, Maria},
  journal={IEEE transactions on medical imaging},
  volume={39},
  number={9},
  pages={2750--2759},
  year={2020},
  publisher={IEEE}
}

@article{xu2023nesvor,
  title={NeSVoR: implicit neural representation for slice-to-volume reconstruction in MRI},
  author={Xu, Junshen and Moyer, Daniel and Gagoski, Borjan and Iglesias, Juan Eugenio and Grant, P Ellen and Golland, Polina and Adalsteinsson, Elfar},
  journal={IEEE transactions on medical imaging},
  volume={42},
  number={6},
  pages={1707--1719},
  year={2023},
  publisher={IEEE}
}

@inproceedings{xu2022svort,
  title={SVoRT: Iterative transformer for slice-to-volume registration in fetal brain MRI},
  author={Xu, Junshen and Moyer, Daniel and Grant, P Ellen and Golland, Polina and Iglesias, Juan Eugenio and Adalsteinsson, Elfar},
  booktitle={International Conference on Medical Image Computing and Computer-Assisted Intervention},
  pages={3--13},
  year={2022},
  organization={Springer}
}

@article{sanchez2024fetmrqc,
  title={FetMRQC: a robust quality control system for multi-centric fetal brain MRI},
  author={Sanchez, Thomas and Esteban, Oscar and Gomez, Yvan and Pron, Alexandre and Koob, M{\'e}riam and Dunet, Vincent and Girard, Nadine and Jakab, Andras and Eixarch, Elisenda and Auzias, Guillaume and others},
  journal={Medical image analysis},
  volume={97},
  pages={103282},
  year={2024},
  publisher={Elsevier}
}

@inproceedings{sanchez2025automatic,
  title={Automatic quality control in multi-centric fetal brain MRI super-resolution reconstruction},
  author={Sanchez, Thomas and Zalevskyi, Vladyslav and Mihailov, Angeline and Mart{\'\i} Juan, Gerard and Eixarch, Elisenda and Jakab, Andras and Dunet, Vincent and Koob, M{\'e}riam and Auzias, Guillaume and Bach Cuadra, Meritxell},
  booktitle={International Workshop on Preterm, Perinatal and Paediatric Image Analysis},
  pages={3--14},
  year={2025},
  organization={Springer}
}

@inproceedings{xu2020semi,
  title={Semi-supervised learning for fetal brain MRI quality assessment with ROI consistency},
  author={Xu, Junshen and Lala, Sayeri and Gagoski, Borjan and Abaci Turk, Esra and Grant, P Ellen and Golland, Polina and Adalsteinsson, Elfar},
  booktitle={International Conference on Medical Image Computing and Computer-Assisted Intervention},
  pages={386--395},
  year={2020},
  organization={Springer}
}

@article{gagoski2022automated,
  title={Automated detection and reacquisition of motion-degraded images in fetal HASTE imaging at 3 T},
  author={Gagoski, Borjan and Xu, Junshen and Wighton, Paul and Tisdall, M Dylan and Frost, Robert and Lo, Wei-Ching and Golland, Polina and van Der Kouwe, Andre and Adalsteinsson, Elfar and Grant, P Ellen},
  journal={Magnetic resonance in medicine},
  volume={87},
  number={4},
  pages={1914--1922},
  year={2022},
  publisher={Wiley Online Library}
}

@article{verdera2025heron,
  title={Heron: High-efficiency real-time motion quantification and re-acquisition for fetal diffusion MRI},
  author={Verdera, Jordina Aviles and Bortolazzi, Antonia and Silva, Sara Neves and Payette, Kelly and Clair, Kamilah St and McElroy, Sarah and Malik, Shaihan and Hajnal, Joseph V and Tomi-Tricot, Raphael and Rutherford, Mary A and others},
  journal={IEEE Transactions on Medical Imaging},
  volume={44},
  number={11},
  pages={4171--4180},
  year={2025},
  publisher={IEEE}
}

@article{hutter2022dynamics,
  title={Dynamics of T2* and deformation in the placenta and myometrium during pre-labour contractions},
  author={Hutter, Jana and Kohli, Vikram and Dellschaft, Neele and Uus, Alena and Story, Lisa and Steinweg, Johannes K and Gowland, Penny and Hajnal, Joseph V and Rutherford, Mary A},
  journal={Scientific Reports},
  volume={12},
  number={1},
  pages={18542},
  year={2022},
  publisher={Nature Publishing Group UK, London}
}

@article{verdera2025comprehensive,
  title={Comprehensive assessment of uterine contractility using a large database of dynamic T2* studies},
  author={Verdera, Jordina Aviles and Payette, Kelly and Hall, Megan and Bradshaw, Charline and Story, Lisa and Bansal, Simi and Waheed, Hadi and Tomi-Tricot, Raphael and Hajnal, Joseph V and Rutherford, Mary and others},
  journal={Placenta},
  year={2025},
  publisher={Elsevier}
}

@article{turk2020placental,
  title={Placental MRI: effect of maternal position and uterine contractions on placental BOLD MRI measurements},
  author={Turk, Esra Abaci and Abulnaga, S Mazdak and Luo, Jie and Stout, Jeffrey N and Feldman, Henry A and Turk, Ata and Gagoski, Borjan and Wald, Lawrence L and Adalsteinsson, Elfar and Roberts, Drucilla J and others},
  journal={Placenta},
  volume={95},
  pages={69--77},
  year={2020},
  publisher={Elsevier}
}

@inproceedings{roth2022towards,
  title={Towards total recall in industrial anomaly detection},
  author={Roth, Karsten and Pemula, Latha and Zepeda, Joaquin and Sch{\"o}lkopf, Bernhard and Brox, Thomas and Gehler, Peter},
  booktitle={2022 IEEE/CVF Conference on Computer Vision and Pattern Recognition (CVPR)},
  pages={14298--14308},
  year={2022},
  organization={IEEE}
}

@article{simeoni2025dinov3,
  title={Dinov3},
  author={Sim{\'e}oni, Oriane and Vo, Huy V and Seitzer, Maximilian and Baldassarre, Federico and Oquab, Maxime and Jose, Cijo and Khalidov, Vasil and Szafraniec, Marc and Yi, Seungeun and Ramamonjisoa, Micha{\"e}l and others},
  journal={arXiv preprint arXiv:2508.10104},
  year={2025}
}

@inproceedings{schlereth2025faster,
  title={Faster, self-supervised super-resolution for anisotropic multi-view MRI using a sparse coordinate loss},
  author={Schlereth, Maja and Schillinger, Moritz and Breininger, Katharina},
  booktitle={International Conference on Medical Image Computing and Computer-Assisted Intervention},
  pages={172--182},
  year={2025},
  organization={Springer}
}

@ARTICLE{SahaBosmaTwilt2024,
  title = {Artificial intelligence and radiologists in prostate cancer detection on MRI (PI-CAI): an international, paired, non-inferiority, confirmatory study},
  journal = {The Lancet Oncology},
  year = {2024},
  issn = {1470-2045},
  volume={25},
  number={7},
  pages={879--887},
  doi = {https://doi.org/10.1016/S1470-2045(24)00220-1},
  author = {Anindo Saha and Joeran S Bosma and Jasper J Twilt and Bram {van Ginneken} and Anders Bjartell and Anwar R Padhani and David Bonekamp and Geert Villeirs and Georg Salomon and Gianluca Giannarini and Jayashree Kalpathy-Cramer and Jelle Barentsz and Klaus H Maier-Hein and Mirabela Rusu and Olivier Rouvière and Roderick {van den Bergh} and Valeria Panebianco and Veeru Kasivisvanathan and Nancy A Obuchowski and Derya Yakar and Mattijs Elschot and Jeroen Veltman and Jurgen J Fütterer and Constant R. Noordman and Ivan Slootweg and Christian Roest and Stefan J. Fransen and Mohammed R.S. Sunoqrot and Tone F. Bathen and Dennis Rouw and Jos Immerzeel and Jeroen Geerdink and Chris {van Run} and Miriam Groeneveld and James Meakin and Ahmet Karagöz and Alexandre Bône and Alexandre Routier and Arnaud Marcoux and Clément Abi-Nader and Cynthia Xinran Li and Dagan Feng and Deniz Alis and Ercan Karaarslan and Euijoon Ahn and François Nicolas and Geoffrey A. Sonn and Indrani Bhattacharya and Jinman Kim and Jun Shi and Hassan Jahanandish and Hong An and Hongyu Kan and Ilkay Oksuz and Liang Qiao and Marc-Michel Rohé and Mert Yergin and Mohamed Khadra and Mustafa E. Şeker and Mustafa S. Kartal and Noëlie Debs and Richard E. Fan and Sara Saunders and Simon J.C. Soerensen and Stefania Moroianu and Sulaiman Vesal and Yuan Yuan and Afsoun Malakoti-Fard and Agnė Mačiūnien and Akira Kawashima and Ana M.M. de M.G. {de Sousa Machadov} and Ana Sofia L. Moreira and Andrea Ponsiglione and Annelies Rappaport and Arnaldo Stanzione and Arturas Ciuvasovas and Baris Turkbey and Bart {de Keyzer} and Bodil G. Pedersen and Bram Eijlers and Christine Chen and Ciabattoni Riccardo and Deniz Alis and Ewout F.W. {Courrech Staal} and Fredrik Jäderling and Fredrik Langkilde and Giacomo Aringhieri and Giorgio Brembilla and Hannah Son and Hans Vanderlelij and Henricus P.J. Raat and Ingrida Pikūnienė and Iva Macova and Ivo Schoots and Iztok Caglic and Jeries P. Zawaideh and Jonas Wallström and Leonardo K. Bittencourt and Misbah Khurram and Moon H. Choi and Naoki Takahashi and Nelly Tan and Paolo N. Franco and Patricia A. Gutierrez and Per Erik Thimansson and Pieter Hanus and Philippe Puech and Philipp R. Rau and Pieter {de Visschere} and Ramette Guillaume and Renato Cuocolo and Ricardo O. Falcão and Rogier S.A. {van Stiphout} and Rossano Girometti and Ruta Briediene and Rūta Grigienė and Samuel Gitau and Samuel Withey and Sangeet Ghai and Tobias Penzkofer and Tristan Barrett and Varaha S. Tammisetti and Vibeke B. Løgager and Vladimír Černý and Wulphert Venderink and Yan M. Law and Young J. Lee and Maarten {de Rooij} and Henkjan Huisman},
}

@inproceedings{zou2024sparse3d,
  title={Sparse3d: Distilling multiview-consistent diffusion for object reconstruction from sparse views},
  author={Zou, Zixin and Cheng, Weihao and Cao, Yan-Pei and Huang, Shi-Sheng and Shan, Ying and Zhang, Song-Hai},
  booktitle={Proceedings of the AAAI conference on artificial intelligence},
  volume={38},
  number={7},
  pages={7900--7908},
  year={2024}
}

@article{wang2024res,
  title={Res-NeuS: Deep residuals and neural implicit surface learning for multi-view reconstruction},
  author={Wang, Wei and Gao, Fengjiao and Shen, Yongliang},
  journal={Sensors},
  volume={24},
  number={3},
  pages={881},
  year={2024},
  publisher={MDPI}
}

@article{duba2025plug,
  title={Plug-and-Play with 2.5 D Artifact Reduction Prior for Fast and Accurate Industrial Computed Tomography Reconstruction},
  author={Duba-Sullivan, Haley and Pramanik, Aniket and Singanallur, Venkatakrishnan and Ziabari, Amirkoushyar},
  journal={Journal of Nondestructive Evaluation},
  volume={44},
  number={4},
  pages={147},
  year={2025},
  publisher={Springer}
}

@article{ahmad2026compensating,
  title={Compensating streak artifacts in sparse-view inline industrial CT for accurate metrology using self-supervised optimization of implicit neural volume representations},
  author={Ahmad, Faizan and Yang, Guangpu and Buchfink, Manuel and Alsaffar, Ammar and Baraka, Ahmed and Liu, Xingyu and Simon, Sven},
  journal={Journal of Nondestructive Evaluation},
  volume={45},
  number={1},
  pages={19},
  year={2026},
  publisher={Springer}
}

@inproceedings{muratbekuly2025research,
  title={Research in structure from motion and multi-view stereo techniques in photogrammetric 3D reconstruction from 2D image sequences.},
  author={Muratbekuly, Beket and Tynymbayev, Sakhybay},
  booktitle={SNE},
  year={2025}
}

@article{sarmah2023survey,
  title={Survey of methods and principles in three-dimensional reconstruction from two-dimensional medical images},
  author={Sarmah, Mriganka and Neelima, Arambam and Singh, Heisnam Rohen},
  journal={Visual computing for industry, biomedicine, and art},
  volume={6},
  number={1},
  pages={15},
  year={2023},
  publisher={Springer}
}

@article{javadi2026video,
  title={Video-based 3D reconstruction: a review of photogrammetry and visual SLAM approaches},
  author={Javadi Moghadam, Ali and Kiani, Abbas and Naeimaei, Reza and Malihi, Shirin and Brilakis, Ioannis},
  journal={Journal of Imaging},
  volume={12},
  number={3},
  pages={128},
  year={2026},
  publisher={MDPI}
}

@article{loris2010nonlinear,
  title={Nonlinear regularization techniques for seismic tomography},
  author={Loris, Ignace and Douma, Huub and Nolet, Guust and Daubechies, Ingrid and Regone, Carl},
  journal={Journal of Computational Physics},
  volume={229},
  number={3},
  pages={890--905},
  year={2010},
  publisher={Elsevier}
}

@article{feng2026respiratory,
  title={Respiratory Motion Management in Abdominal MRI: Revisiting the Gap Between Technical Advances and Clinical Translation},
  author={Feng, Li and Chandarana, Hersh},
  journal={Magnetic resonance in medicine},
  year={2026},
  publisher={Wiley Online Library}
}

@article{wang2026learning,
  title={Learning-based multi-view stereo: A survey},
  author={Wang, Fangjinhua and Zhu, Qingtian and Chang, Di and Gao, Quankai and Han, Junlin and Zhang, Tong and Hartley, Richard and Pollefeys, Marc},
  journal={IEEE Transactions on Pattern Analysis and Machine Intelligence},
  year={2026},
  publisher={IEEE}
}

@article{uus2022automated,
  title={Automated 3D reconstruction of the fetal thorax in the standard atlas space from motion-corrupted MRI stacks for 21--36 weeks GA range},
  author={Uus, Alena U and Grigorescu, Irina and van Poppel, Milou PM and Steinweg, Johannes K and Roberts, Thomas A and Rutherford, Mary A and Hajnal, Joseph V and Lloyd, David FA and Pushparajah, Kuberan and Deprez, Maria},
  journal={Medical image analysis},
  volume={80},
  pages={102484},
  year={2022},
  publisher={Elsevier}
}

@article{moreira2018clinical,
  title={Clinical study of fetal neurobehavior by the KANET test},
  author={Moreira Neto, Raul and Porovic, Selma},
  journal={Journal of perinatal medicine},
  volume={46},
  number={6},
  pages={631--639},
  year={2018},
  publisher={De Gruyter}
}

@article{beausejour2024innovations,
  title={Innovations in the assessment of skeletal muscle health: A glimpse into the future},
  author={Beausejour, Jonathan P and Knowles, Kevan S and Wilson, Abigail T and Mangum, L Colby and Hill, Ethan C and Hanney, William J and Wells, Adam J and Fukuda, David H and Stout, JeffreyR and Stock, Matt S},
  journal={International Journal of Sports Medicine},
  volume={45},
  number={09},
  pages={659--671},
  year={2024},
  publisher={Georg Thieme Verlag}
}

@article{elad2020biomechanics,
  title={Biomechanics of early life in the female reproductive tract},
  author={Elad, David and Jaffa, Ariel J and Grisaru, Dan},
  journal={Physiology},
  volume={35},
  number={2},
  pages={134--143},
  year={2020},
  publisher={American Physiological Society, Bethesda, MD}
}

@article{hameed2025esr,
  title={ESR Essentials: diagnosis and assessment of treatment response in patients with luminal Crohn’s disease—practice recommendations by the European Society of Gastrointestinal and Abdominal Radiology},
  author={Hameed, Maira and De Kock, Isabelle and Stoker, Jaap and Taylor, Stuart A},
  journal={European radiology},
  volume={35},
  number={12},
  pages={7484--7496},
  year={2025},
  publisher={Springer}
}

@article{weinreb2016pi,
  title={PI-RADS prostate imaging--reporting and data system: 2015, version 2},
  author={Weinreb, Jeffrey C and Barentsz, Jelle O and Choyke, Peter L and Cornud, Francois and Haider, Masoom A and Macura, Katarzyna J and Margolis, Daniel and Schnall, Mitchell D and Shtern, Faina and Tempany, Clare M and others},
  journal={European urology},
  volume={69},
  number={1},
  pages={16--40},
  year={2016},
  publisher={Elsevier}
}

@article{coussot2020progress,
  title={Progress in rheology and hydrodynamics allowed by NMR or MRI techniques},
  author={Coussot, Philippe},
  journal={Experiments in Fluids},
  volume={61},
  number={9},
  pages={207},
  year={2020},
  publisher={Springer}
}

@article{alves2021detection,
  title={Detection and imaging of damages and defects in fibre-reinforced composites by magnetic resonance technique},
  author={Alves, Carine L and Oliveira, Janete S and Tannus, Alberto and Tarpani, Alessandra Cristina Soares P and Tarpani, Jose R},
  journal={Materials},
  volume={14},
  number={4},
  pages={977},
  year={2021},
  publisher={MDPI}
}

@inproceedings{dai2008motion,
  title={Motion from blur},
  author={Dai, Shengyang and Wu, Ying},
  booktitle={2008 IEEE Conference on Computer Vision and Pattern Recognition},
  pages={1--8},
  year={2008},
  organization={IEEE}
}

@article{savarese20073d,
  title={3d reconstruction by shadow carving: Theory and practical evaluation},
  author={Savarese, Silvio and Andreetto, Marco and Rushmeier, Holly and Bernardini, Fausto and Perona, Pietro},
  journal={International journal of computer vision},
  volume={71},
  number={3},
  pages={305--336},
  year={2007},
  publisher={Springer}
}

@article{pentland1987new,
  title={A new sense for depth of field},
  author={Pentland, Alex Paul},
  journal={IEEE transactions on pattern analysis and machine intelligence},
  year={1987},
  publisher={IEEE}
}

@article{bouman2017turning,
  title={Turning corners into cameras: Principles and methods},
  author={Bouman, Katherine L and Ye, Vickie and Yedidia, Adam B and Durand, Fr{\'e}do and Wornell, Gregory W and Torralba, Antonio and Freeman, William T},
  year={2017},
  publisher={Institute of Electrical and Electronics Engineers (IEEE)}
}

@inproceedings{ait2006simultaneous,
  title={Simultaneous object pose and velocity computation using a single view from a rolling shutter camera},
  author={Ait-Aider, Omar and Andreff, Nicolas and Lavest, Jean Marc and Martinet, Philippe},
  booktitle={European Conference on Computer Vision},
  pages={56--68},
  year={2006},
  organization={Springer}
}

@article{ji2023fetal,
  title={Fetal behavior during MRI changes with age and relates to network dynamics},
  author={Ji, Lanxin and Majbri, Amyn and Hendrix, Cassandra L and Thomason, Moriah E},
  journal={Human Brain Mapping},
  volume={44},
  number={4},
  pages={1683--1694},
  year={2023},
  publisher={Wiley Online Library}
}
}

\clearpage
\setcounter{page}{1}
\maketitlesupplementary
\appendix
\setcounter{table}{0}\setcounter{figure}{0}
\counterwithin*{table}{section}\counterwithin*{figure}{section}
\renewcommand{\thetable}{\thesection\arabic{table}}
\renewcommand{\thefigure}{\thesection\arabic{figure}}

\section{Overview of the Experiments}\label{app:overview}
Table~\ref{tab:experiments} lists every experiment of the paper with the question it answers, its data, setup and evaluation.
\begin{table*}[p]
\centering\scriptsize
\caption{Overview of the experiments. For each experiment, the question it answers, the data, the setup with its parameters and how it is evaluated. U: uterus, F: fetal brain, P: prostate. Unless stated otherwise, all fits use the one configuration of Sec.~\ref{sec:optim} (three stages of 1000 iterations, $A$ on 16 / 8 / 4\,mm grids, $T_v$ on 8 / 4\,mm grids, deformation on a 5\,mm grid, $\lambda_T = 0.2$, $c = 0.05$). Every setting was fixed before the evaluation it is reported on, and no reconstruction setting was chosen on P.}
\label{tab:experiments}
\begin{tabularx}{\textwidth}{>{\raggedright\arraybackslash}p{2.6cm} >{\raggedright\arraybackslash}p{2.9cm} >{\raggedright\arraybackslash}X >{\raggedright\arraybackslash}p{3.4cm}}
\toprule
Experiment (question) & Data & Setup and parameters & Evaluation \\
\midrule
\multicolumn{4}{l}{\emph{Reconstruction}} \\
Held-out stack prediction (Sec.~\ref{sec:recon}) \newline Does the volume predict an acquisition it has not seen? &
U: all 123 patients, 369 cases (cases scored where every method returned a volume) \newline F: 83 fetuses = 65 healthy test fetuses + 18 with pathology, 249 cases (one stack without a usable brain mask) \newline P: the 85 healthy test patients, 255 cases &
Each stack $h$ left out in turn, all methods reconstruct from the remaining stacks. $A$ frozen, stack $h$ registered anew to every volume (rigid, non-rigid, gain, offset and smooth bias, 400 iterations, no transient) and simulated through its slice profile. Methods: ours, SVRTK (template by fixed rule), NeSVoR (SVoRT for F, stack registration for U and P), mean of the other stacks with and without our motion correction. &
NCC, SSIM, PSNR in a common region (U, P: organ + 10\,mm, F: brain mask), also after a common in-plane low-pass ($\sigma$ = 1, 2, 4\,mm) and per frequency band, median over cases, per-patient Wilcoxon test, better if median difference favors a method and $p < 0.05$. \\
Seen-stack consistency (Sec.~\ref{sec:recon}) \newline How closely does the volume reproduce its own inputs? &
The same subjects as above: U 123, F 83, P 85 &
Same registration and simulation as above, applied to every stack the volume was reconstructed from. &
As above. Interpreted together with the held-out result, since keeping single-stack changes raises this score. \\
\midrule
\multicolumn{4}{l}{\emph{Separation}} \\
Digital phantom (Sec.~\ref{sec:phantom}) \newline Are static and transient changes attributed to the right component? &
U, P: 20 patients each, seeded draw (U: among the patients with an SVRTK reconstruction, without the patients of the benchmarks used for the configuration, P: among the 85 test patients) &
Ground truth $\mathrm{GT}$: SVRTK reconstruction. Simulated stacks with motion $\pm3^\circ$, $\pm4$\,mm, smooth deformation 1--3\,mm, trapezoidal slice profile, in-plane blur, per-slice jitter 0.3\,mm, polynomial bias $\pm10\%$, Rician noise. Per patient one case without and two with changes: two transients (one stack each, spheres $r = 5$, 8 or 12\,mm, contrast $\pm40\%$), one compression up to 3\,mm (one stack), one static change (all stacks). &
NCC and SSIM of $A$ against ground truth, motion error (mm), share of each change in $A$ and in its $T_v$ (projection onto the inserted change, change isolated by subtracting the unchanged case). \\
Ablation (Sec.~\ref{sec:ablation}) \newline Which parts of the model matter? &
U, F: 20 patients each, seeded draw from the training split \newline U phantom: the 20 phantoms above and a harder variant \newline F: the corruption benchmark below &
One change at a time: rigid only, no $T_v$, no robust reweighting, $T_v$ from the non-rigid stage, reference stack instead of centered gauge, no slice profile, $A$ down to 2\,mm, $T_v$ grid 4 / 2\,mm. Our configuration refitted with another seed as reference for the refit variability. Hard phantom: noise $\times 2$, jitter 1\,mm, texture 5\%. &
Median per-patient held-out NCC difference to ours, phantom share of transients in $T_v$ and $A$, SSIM of $A$, flagging AUROC. \\
\midrule
\multicolumn{4}{l}{\emph{Transients as a signal}} \\
Corruption benchmark (Sec.~\ref{sec:qc}) \newline Does $T_v$ find corrupted slices? &
F: 20 healthy fetuses, seeded draw from the training split, 180 corrupted slices &
Three slices per stack corrupted by signal dropout, in-plane motion with ghosting or displaced tissue, then refitted. Variants: all three stacks, two fitted stacks, a new stack checked against the anatomy of two earlier ones. &
AUROC of the transient strength $s$ for corrupted against clean slices, per corruption type. Recovery of the original content: NCC and PSNR of the slice predicted from $A$. \\
Expert slice ratings (Sec.~\ref{sec:qc}) \newline Does $T_v$ agree with expert ratings? &
F: 30 healthy fetuses, seeded draw among those with a brain mask on every stack and not in the corruption benchmark, 90 stacks, 65 clearly and 45 mildly corrupted, 1731 clean slices &
Two experts, each fetus rated by one of them, blinded to the model. Flag threshold: 95th percentile of $s$ over all slices of the 239 examinations in no slice-quality evaluation (also 90th and 98th). &
AUROC with 95\% bootstrap CI over stacks, sensitivity and false-positive rate at each threshold. \\
Timing of flagged slices (Sec.~\ref{sec:qc}) \newline Do flagged slices form events in time? &
F: all 289 fetuses, the 866 interleaved stacks of 867 &
Flagged slices at the 95th percentile, neighbors in acquisition order against neighbors in spatial order. &
Share with a flagged neighbor, two permutation tests (random placement, circular shift in spatial order). \\
Flagged slices over gestation (Sec.~\ref{sec:qc}) \newline Does the share of flagged slices change with gestational age? &
F: 107 healthy fetuses of a second cohort (17.1--39.3 weeks, first examination per fetus, those of 165 additional fits with a documented gestational age), pooled with the 271 healthy and 15 affected fetuses of the main dataset with a documented gestational age &
Unchanged configuration. Per fetus: share of flagged slices (threshold above) and displacement of the brain between the stacks. Brain masks that failed in a stack with strong motion ignored, the stack kept. &
Spearman correlation with gestational age, partial for protocol and slice thickness (second cohort) or cohort (pooled), robustness without fetuses with a failed mask. \\
\bottomrule
\end{tabularx}
\end{table*}
\begin{table*}[p]
\addtocounter{table}{-1}
\centering\scriptsize
\caption{Overview of the experiments (continued).}
\begin{tabularx}{\textwidth}{>{\raggedright\arraybackslash}p{2.6cm} >{\raggedright\arraybackslash}p{2.9cm} >{\raggedright\arraybackslash}X >{\raggedright\arraybackslash}p{3.4cm}}
\toprule
\multicolumn{4}{l}{\emph{Static pathology}} \\
Lesion insertion benchmark (Sec.~\ref{sec:anomalyexp}) \newline Which input should the detector use? &
U, F, P: 20 healthy test subjects each (U: all 20 test patients, F, P: seeded draw from the test split), 2 cases per subject &
Per case three static lesions (all stacks, contrast $\pm40\%$, $r = 3$, 5, 8 or 12\,mm, inside the organ or at its boundary, $\geq 8$\,mm apart) and two transient distractors (one stack, contrast $\pm40\%$, $r = 5$ or 8\,mm, $\geq 4$\,mm from other changes), rendered through each stack's fitted motion, slice profile, gain and bias, then refitted. Candidates (U, F, P): $A$, mean of the motion-corrected stacks, that mean weighted by $T_v$ and $\det J$, second-largest stack, $A$ with 2\,mm grid. &
Pre-registered criterion: mean over strata (contrast $\times$ size $\times$ location) of the median voxel AUROC, the simplest candidate within 0.01 of the best wins. Share of distractor voxels with $S \geq 2$. \\
Real lesion detection (Sec.~\ref{sec:anomalyexp}) \newline How well does each input localize real lesions? &
U: the 37 patients with radiologist masks in every stack showing the lesion \newline F: the 18 fetuses with pathology \newline P: the 47 patients with clinically significant cancer and an expert delineation (axial stack) &
Frozen detector: PatchCore on DINOv3 tokens at two scales (8 and 16\,mm), PCA and memory from the healthy training subjects (U: 40, F: 196, P: 118, i.e.\ training split without the 10 calibration subjects), tail calibration on these 10 held-out healthy subjects, $S = -\log_{10} p$, maximum over scales, Gaussian smoothing 1.5\,mm. Evaluated once, U also per finding and without deep infiltrating endometriosis (declared from the radiologist's rating of label reliability). P: $A$ with the frozen detector, the other inputs with banks rebuilt for the selection. &
Per-patient voxel AUROC of lesion voxels against all others in organ + 40\,mm seen by $\geq 2$ stacks, excluding 3\,mm around each lesion, lesions in $\geq 1$ and $\geq 2$ stacks (P: $\geq 1$), by polarity. False positives: share of healthy voxels with $S \geq 2$. \\
Detector on SVRTK and NeSVoR (Sec.~\ref{sec:anomalyexp}) \newline Is $A$ a better input than SVR volumes? &
Benchmark: U 37 cases (SVRTK failed on 3 of 40), F 40, P 40 \newline Real lesions: U 37 patients, F 18 fetuses &  benchmark cases with all three inputs
Same detector, PCA and memory rebuilt per input from that input's healthy training subjects, volumes registered rigidly to $A$, one common region. &
Benchmark criterion as above, $A$ better if ahead by more than 0.02. Real lesions: per-patient AUROC, Wilcoxon test. \\
Normative model (Sec.~\ref{sec:anomalyexp}) \newline Memory bank or normalizing flow? &
U, F: insertion benchmark &
Same tokens and held-out calibration, nearest-neighbor memory against a normalizing flow. &
Benchmark criterion, $-\log p$ on training against unseen healthy subjects. \\
Same-cohort fetal reference (Sec.~\ref{sec:anomalyexp}) \newline Does the cohort shift inflate the fetal results? &
F: 165 healthy control fetuses of the cohort and protocol of the fetuses with pathology (25\% test, seeded, one without coverage skipped), the 18 fetuses with pathology &
Frozen detector, PCA, memory and calibration rebuilt from the training controls of that cohort, inputs $A$ and the mean of the motion-corrected stacks. Declared before running. &
Lesion AUROC of the 18 fetuses against the original reference, patient-level AUROC against the held-out controls of the same cohort. \\
Single-vendor reference (Sec.~\ref{sec:anomalyexp}, exploratory) \newline Does scanner heterogeneity explain the prostate result? &
P: Siemens 31 / 69 / 102, Philips 16 / 16 / 26 (diseased / test / training) &
PCA, memory and calibration from the healthy training subjects of one vendor, declared before running. &
As for real lesions, plus gland volume as a confound. \\
Transient location (Sec.~\ref{sec:anomalyexp}) \newline Where does the transient signal lie? &
U: 60 fits, P: 60 fits &
Voxels with $|T_v| > 0.08$, in the organ + 10\,mm against a shell 10--40\,mm around it, covered by $\geq 2$ stacks. &
Density outside against inside the organ. \\
\bottomrule
\end{tabularx}
\end{table*}
\clearpage

\section{Held-out Stack Prediction}\label{app:heldout}
\begin{figure*}[!h]
\centering
\includegraphics[width=\textwidth]{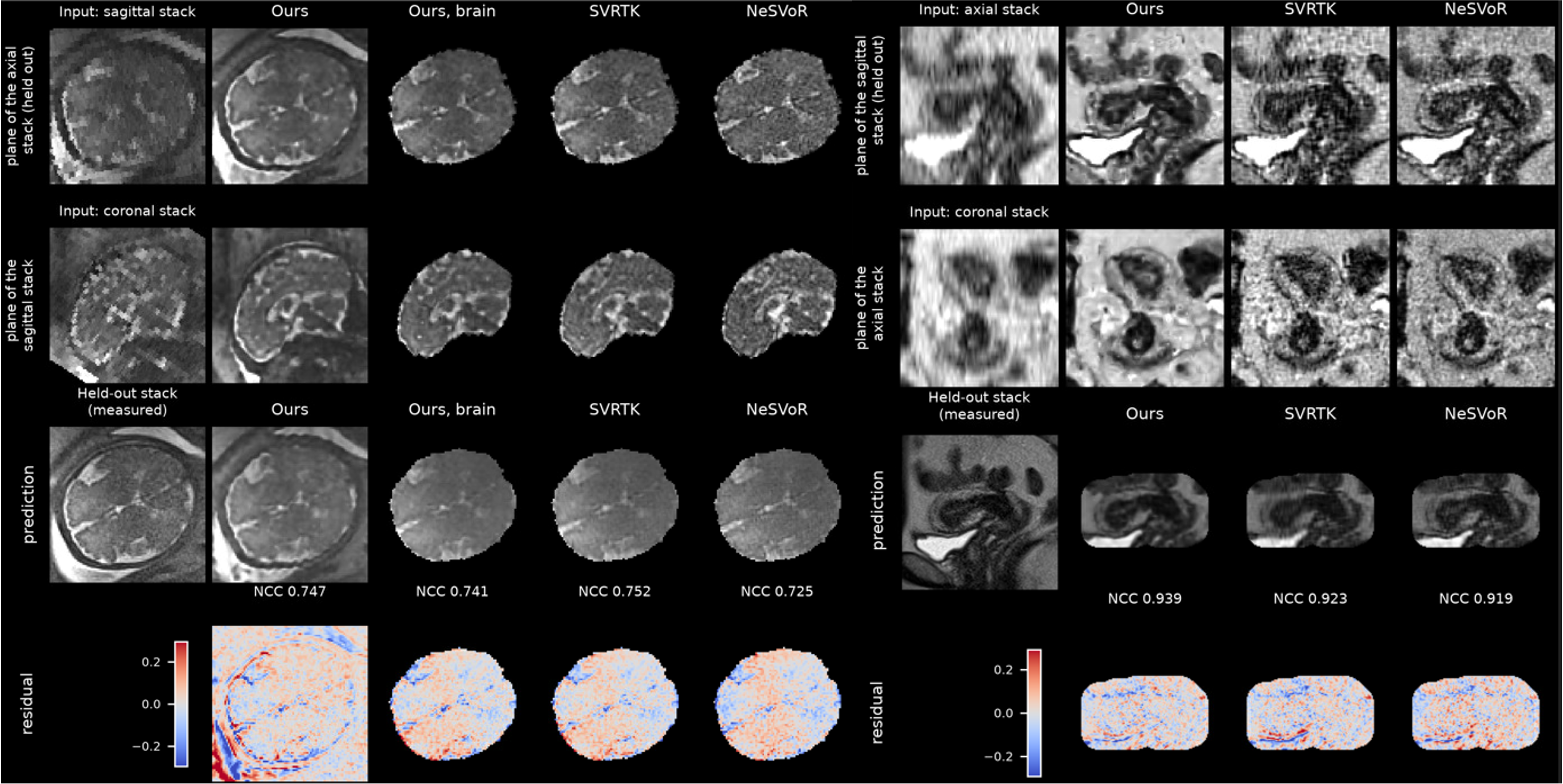}
\caption{Reconstructions and held-out prediction for a fetal brain (left) and a uterus (right), each the subject with the median per-patient difference between ours and SVRTK in the held-out evaluation. Rows 1--2: the reconstructions from all three stacks in the slice plane of the held-out stack (row 1) and of a second stack (row 2), next to a fitted stack resampled into that plane, which displays its 4.5\,mm slice thickness. The fetal stacks are tilted with respect to the scanner and named after the fetal head. SVRTK and NeSVoR are rigidly registered to our grid. Rows 3--4: the held-out stack as measured, its prediction by each method from the other two stacks after the identical registration, with the NCC over the scored region, and the residual on one color scale. Predictions are displayed with a least-squares gain and offset to the measured stack, after the NCC was computed. SVRTK, NeSVoR and ours, brain, restricted to the brain reconstruct only the brain and are displayed inside it. Our full reconstruction also predicts the maternal tissue and is presented over the whole field of view, with the NCC computed in the brain for all methods.}
\label{fig:recon}
\end{figure*}

Fig.~\ref{fig:recon} depicts, for a fetal brain and a uterus, the reconstructions of all methods from identical inputs and their prediction of a stack none of them saw (Sec.~\ref{sec:recon}).

\section{Anomaly Detection on the Static Anatomy}\label{sec:anomalyexp}
As a downstream use of the decomposition, we screen its static anatomy for pathology with a frozen detector. We use a memory-bank detector~\cite{roth2022towards} on frozen self-supervised features to evaluate $A$ as a candidate input for screening static pathology. Patch tokens of DINOv3~\cite{simeoni2025dinov3} at two scales, computed on $A$ after a per-subject intensity normalization and reduced by a PCA fitted on healthy subjects, are scored by their distance to the nearest token of a bank built from the same healthy training subjects. For each scale, the distance of a patch to its nearest healthy token is converted into an empirical tail probability $p$, the fraction of patches of ten held-out healthy subjects with at least this distance. The two scales are combined by the smaller $p$, giving the map $S = -\log_{10} p$ on the grid of $A$ ($S = 2$ for $p = 0.01$). The detector does not depend on how its input volume was reconstructed. Any volume on a common grid can be scored by building the bank from the same kind of volume of the healthy training subjects, which makes the choice of input a controlled comparison. Detection maps are scored per subject by the voxel-level AUROC of lesion voxels against all other voxels (0.5 chance, 1 perfect) in a region covering the organ and 40\,mm around it where at least two stacks see the tissue, excluding a 3\,mm margin around each lesion. We report lesions marked in at least one and two stacks, separately by polarity, and as false positives the share of healthy tissue with $S \ge 2$, \ie $p \le 0.01$ (lower is better).

\emph{Lesion insertion benchmark.} Lesions with known position are inserted into real acquired stacks from healthy subjects not part of model training to choose and compare the inputs of the anomaly detector. The thus changed anatomy is rendered into the stacks through each stack's fitted motion, slice profile, gain and bias and the subject is then refitted from these stacks. Both static lesions present in every stack and transient distractors of the same kind present in one stack only are inserted. This mirrors the clinical pitfall of uterine imaging, where a contraction seen by one stack mimics a lesion. All changes are smooth spheres with a contrast of $\pm 40\%$. Lesions have radii of 3, 5, 8 and 12\,mm, from about one slice thickness to several slices, placed inside the organ or at its boundary. Distractors take the middle radii of 5 and 8\,mm, differing hence only in being seen by one stack. Each case contains three lesions and two distractors, with two cases on each of 20 healthy subjects per dataset. For scoring, the known lesions are mapped onto the grid of each refit through its fitted motion.

\emph{Expert annotations.} Uterine lesions (adenomyosis, endometriosis and myomas) were segmented in every stack of 37 patients by a specialist radiology resident, fetal brain lesions (ventriculomegaly, pseudocysts and scalp tumors) were segmented by one observer, and for the prostate the expert tumor delineations released with PI-CAI on the axial stack were taken. The expert labels are used only to evaluate methods whose settings had been fixed beforehand. In addition, corrupted slices in the fetal stacks were each rated by one of two experts blinded to the model (Sec.~\ref{sec:qc}).

\emph{Selection on synthetic lesions.} The input to the detector was chosen on the lesion insertion benchmark (Sec.~\ref{sec:protocols}). Candidates were $A$, the mean of the motion-corrected stacks, that mean weighted down where $T_v$ or the volume change is large, the second-largest per-stack score, and $A$ of a fit with a 2\,mm finest grid. By the pre-registered rule the winner is $A$ for the uterus, the mean of the motion-corrected stacks for the fetal brain and the weighted mean for the prostate. No candidate flagged the inserted transients (share of transient-distractor voxels with $S \ge 2$: 0).

\emph{Real lesions.} Table~\ref{tab:anomaly} reports every input once on the expert lesion masks, as per-patient lesion AUROC within each subject. On the uterus (37 patients), $A$ reaches 0.74 (lesions marked in at least one stack) and 0.76 (at least two stacks). No other input is better per patient. The mean of the motion-corrected stacks has a higher median (0.77 / 0.79) but is better in only 19 of 37 patients ($p = 0.46$), and the second-largest per-stack score and the 2\,mm grid are worse ($p = 0.006$ and $p = 0.009$). Detection is highest for endometriomas (0.90), deep infiltrating endometriosis (0.84) and myomas (0.79), and lowest for diffuse adenomyosis (0.63). By polarity, hyperintense lesions reach 0.86 against 0.68 and 0.74 for hypo- and isointense lesions. Weighting the map down where the transient field or the volume change is large removes 97\% of the false positives on healthy patients at little cost on the uterus (0.72 against 0.74, $p = 0.93$) but loses real lesions on the fetal brain (0.69 against 0.89), where part of a lesion's signal can also end up in $T_v$ at 4.5\,mm slices. On the fetal brain, the selected mean of the motion-corrected stacks beats $A$ in 17 of 18 patients ($p < 0.001$). On the prostate, the frozen detector on $A$ does not localize clinically significant cancer (0.38, 47 patients). The selected weighted mean reaches 0.63 and beats $A$ under the same newly built banks (0.44, better in 37 of 47 patients, $p < 10^{-4}$). The weight alone, without any detector, already reaches 0.62, however, and the detector on the unweighted mean stays below chance (0.47). This gain therefore reflects where the fit places transients and volume changes, which is partly the gland itself, not detection.

\begin{table}[!t]
\centering\small
\caption{Lesion detection on expert lesion masks (median per-patient lesion AUROC, $\ge 1$ / $\ge 2$ stacks). Prostate: masks on the axial stack only, $A$ with the frozen detector, the other inputs with banks built anew for the selection, under which $A$ reaches 0.44. Healthy FP: share of voxels with $S \ge 2$ on healthy test patients.}
\label{tab:anomaly}
\resizebox{\columnwidth}{!}{
\begin{tabular}{lcccc}
\toprule
Input & Uterus (37) & Fetal (18) & Prostate (47) & Healthy FP (uterus) \\
\midrule
$A$ & 0.74 / 0.76 & 0.88 / 0.93 & 0.38 / -- & 0.34\% \\
Mean of motion-corrected stacks & 0.77 / 0.79 & 0.89 / 0.95 & 0.47 / -- & 0.23\% \\
Mean, weighted by $T_v$ and $\det J$ & 0.72 / 0.74 & 0.69 / 0.63 & 0.63 / -- & 0.01\% \\
Second-largest stack & 0.71 / 0.67 & 0.84 / 0.90 & 0.45 / -- & 0.02\% \\
$A$, 2\,mm grid & 0.70 / 0.68 & 0.88 / 0.92 & 0.42 / -- & 0.06\% \\
\bottomrule
\end{tabular}
}
\end{table}

\emph{Comparison with SVR volumes.} The same frozen detector, with its memory of healthy tokens built from that input's own healthy training patients, was applied to $A$ and to the SVRTK and NeSVoR volumes from identical inputs, registered rigidly to $A$ and restricted to one common region. On the uterine benchmark (37 cases with all three inputs), $A$ reaches a criterion of 0.70 against 0.65 for SVRTK and 0.67 for NeSVoR, better per case in 30 and 27 of 37 cases ($p < 10^{-4}$ and $p = 0.002$), which meets the pre-registered rule for both. On healthy hosts $A$ has more voxels with $S \ge 2$ (0.51\% against 0.09\% and 0.04\%), and no input flagged the inserted transients, so the advantage on this benchmark lies in detecting lesions, not suppressing transients. On the expert lesion masks of all 37 patients, $A$ reaches a median per-patient lesion AUROC of 0.76 ($\ge 1$ stack) and 0.77 ($\ge 2$ stacks), against 0.71 and 0.64 for SVRTK and 0.71 and 0.67 for NeSVoR, better per patient in 27 and 28 of 37 patients ($p < 10^{-4}$ and $p < 10^{-3}$). Without deep infiltrating endometriosis, whose masks are least reliable per voxel, the gap grows (29 patients, 0.74 against 0.62 and 0.65, better in 22 and 23 of 29). On the fetal benchmark (40 cases), $A$ reaches a criterion of 0.66 against 0.64 for SVRTK and for NeSVoR, a difference below the pre-registered margin of 0.02, so the rule finds no difference. On the real fetal lesions (18 fetuses) the three inputs reach the same AUROC (0.85 / 0.92 for $A$, 0.85 / 0.92 for SVRTK and 0.86 / 0.91 for NeSVoR, $\ge 1$ / $\ge 2$ stacks, median per-patient difference 0.000, $p = 0.97$ and $p = 0.70$). On the prostate benchmark (40 cases, same design), $A$ reaches a criterion of 0.55 against 0.43 for SVRTK and 0.47 for NeSVoR, better per case in 38 and 30 of 40 cases ($p < 10^{-9}$ and $p = 0.002$), which meets the rule for both, with fewer voxels with $S \ge 2$ on healthy hosts (0.10\% against 0.34\% and 0.29\%). The absolute level is low, however, and on the real expert masks of clinically significant cancer the frozen detector on $A$ does not localize lesions (median per-patient AUROC 0.38, 47 patients). A reference from one vendor alone does not change this (0.37 on the Siemens patients).

\noindent \emph{Normative model: nearest neighbors vs.\ normalizing flow.} On the same tokens and with the same held-out calibration, the memory bank outperforms a normalizing flow on synthetic lesions (uterus 0.689 vs.\ 0.604, fetal 0.625 vs.\ 0.527). A flow on 16\,mm tokens memorized its training patients when calibrated on them (median $-\log p$ 0.60 on training against 2.88 on unseen healthy). Held-out calibration and early stopping removed this gap, but with few healthy subjects the nearest-neighbor model remains more robust.

\noindent \emph{Limits.} The healthy training fetuses and the fetuses with lesions differ in cohort, protocol and gestational age (35.9--40.9 against 26.3--38.9 weeks). A patient-level score would therefore mostly separate the cohorts, so all lesion results are computed within each subject. With a healthy reference from 165 control fetuses of the cohort and protocol of the fetuses with lesions instead, the localization holds (0.88 / 0.92 for $A$, 0.86 / 0.91 for the mean), while the patient-level score no longer separates fetuses with and without pathology (AUROC 0.61 for $A$, 0.57 for the mean). The within-subject results therefore do not stem from the cohort shift. On the prostate, the detector does not localize clinically significant cancer (AUROC 0.38). In T2-weighted images alone these lesions are often subtle, and clinical reading relies on diffusion-weighted imaging for the peripheral zone, where most of them arise~\cite{weinreb2016pi}. In addition, the transient signal lies mostly in the surrounding pelvis, 1.7 times denser there than in and near the gland (1.1 for the uterus).

\section{Ablation}\label{sec:ablation}
The components are designed for the separation, so we measure them where it is known, on the phantom, and on the task that uses $T_v$, slice flagging. Held-out stacks check that none of them costs fidelity (Table~\ref{tab:ablation}). They are predicted from $A$ alone, so they cannot show the separation itself.
\emph{Separation.} Without $T_v$, 0.27 of each transient enters $A$, close to the third expected from averaging three stacks, against 0.08 with it. Without robust reweighting 0.17 remain, without the slice profile 0.15, and with a transient grid finer than the slices (4/2 instead of 8/4\,mm) 0.14. Static changes stay in $A$ in every variant. On a harder phantom (doubled noise, 1\,mm jitter, fine texture), every ablated variant leaks more of a transient into $A$ than ours (0.18--0.31 against 0.15), while $A$ is slightly closer to the ground truth without $T_v$ (NCC 0.945 against 0.933). The slice profile and the 4\,mm limit of $A$ also keep $A$ closer to the ground truth (SSIM 0.899 against 0.887 without profile and 0.854 at 2\,mm), since they keep $A$ from absorbing the blur of each stack and from fitting noise that the slices cannot constrain. Non-rigid motion reduces the motion error from 0.76 to 0.28\,mm. These shares are measured on synthetic transients, so they show where a change ends up, not how often such changes occur in patients.
\emph{Slice flagging.} On the fetal corruption benchmark, the AUROC of the transient strength drops from 0.90 to 0.87 without the slice profile (ghosting 0.79 to 0.71) and to 0.82 with $T_v$ already in the non-rigid stage (ghosting 0.62). Robust reweighting leaves flagging unchanged (0.90), its role is to keep transients out of $A$.
\emph{Fidelity.} Refitting with another random seed changes the held-out NCC by less than 0.001 on the uterus ($p = 0.37$) and by 0.002 on the fetal brain, and the phantom shares by at most 0.01, so differences of this size are within refitting. No variant changes the held-out NCC by more than 0.02, and no component trades fidelity for separation. The largest decreases come from omitting the slice profile (0.004 / 0.015, uterus / fetal), the 2\,mm level of $A$ (0.008 / 0.010) and fitting $T_v$ already in the non-rigid stage on the fetal brain (0.018), consistent with $T_v$ absorbing misalignment while the motion is still estimated. Without $T_v$ or reweighting the held-out NCC rises slightly (at most 0.007), consistent with $T_v$ also taking up detail that only one stack resolves sharply. A reference stack instead of the centered frame predicts almost as well ($-0.001$ / $-0.003$).
\begin{table}[!t]
\centering\small
\setlength{\tabcolsep}{2.5pt}
\caption{Ablation. Separation on the phantom (uterus, 20 patients): median share of an inserted transient in the transient field of its stack ($T_v$) and in $A$, and SSIM of $A$ against the ground truth. Static changes stay in $A$ in every variant. Flagging: AUROC of the transient strength for corrupted slices on the fetal corruption benchmark (20 fetuses, 180 corrupted slices). Fidelity: median per-patient change of the held-out NCC against ours (0.907 uterus, 0.768 fetal, 20 patients per dataset, every stack held out in turn, scored as in Table~\ref{tab:recon}). $^\circ$: not significant (Wilcoxon $p \ge 0.05$). --: no transient field, not run, or not defined because the phantom's ground truth lies in the centered frame.}
\label{tab:ablation}
\resizebox{\columnwidth}{!}{
\begin{tabular}{lccccc}
\toprule
 & \multicolumn{2}{c}{Separation (phantom)} & Flagging & \multicolumn{2}{c}{Fidelity, $\Delta$NCC} \\
\cmidrule(lr){2-3}\cmidrule(lr){4-4}\cmidrule(lr){5-6}
 & $T_v$ $\uparrow$ / $A$ $\downarrow$ & SSIM $\uparrow$ & AUROC $\uparrow$ & Uterus & Fetal \\
\midrule
Ours & 0.66 / 0.08 & 0.899 & 0.90 & -- & -- \\
Ours, other seed & 0.65 / 0.09 & 0.898 & -- & $0.000^\circ$ & $+0.002$ \\
\midrule
\multicolumn{6}{l}{\emph{Transient field}} \\
No $T_v$, rigid only & -- / 0.23 & 0.873 & -- & $0.000^\circ$ & $-0.001^\circ$ \\
No $T_v$, non-rigid & -- / 0.27 & 0.882 & -- & $+0.004$ & $+0.007$ \\
W/o robust reweighting & 0.57 / 0.17 & 0.899 & 0.90 & $+0.003$ & $+0.006$ \\
$T_v$ from the non-rigid stage & 0.67 / 0.08 & 0.898 & 0.82 & $-0.002$ & $-0.018$ \\
$T_v$ grid 4 / 2\,mm & 0.63 / 0.14 & 0.892 & -- & $0.000^\circ$ & $0.000^\circ$ \\
\midrule
\multicolumn{6}{l}{\emph{Anatomy and geometry}} \\
W/o slice profile & 0.55 / 0.15 & 0.887 & 0.87 & $-0.004$ & $-0.015$ \\
$A$ down to 2\,mm & 0.62 / 0.09 & 0.854 & -- & $-0.008$ & $-0.010$ \\
Reference stack & -- & -- & -- & $-0.001$ & $-0.003^\circ$ \\
\bottomrule
\end{tabular}
}
\end{table}

\end{document}